\documentclass[11pt]{article}
\usepackage[utf8]{inputenc}
\usepackage[makeroom]{cancel}
\usepackage{amsmath}
\usepackage[22pt, Times New Roman]{}
\usepackage{subfiles}
\newcommand{\norm}[1]{\| #1 \|}

\usepackage{setspace}
\usepackage{caption}
\usepackage{float}
\usepackage{wrapfig}
\usepackage{blkarray}
\usepackage{subcaption}
\usepackage{indentfirst}
\usepackage[english]{babel}
\usepackage{url}
\usepackage[nottoc]{tocbibind}

\usepackage{geometry}
\usepackage{graphicx}
\usepackage{graphicx}
\usepackage[Times New Roman, 12pt]{}
\graphicspath{{./images/} }
\usepackage{tikz}
\usetikzlibrary{shapes.geometric, arrows}
\tikzstyle{startstop} = [rectangle, rounded corners, minimum width=3cm, minimum height=0.5cm,text centered, text width=8cm, draw=black, fill=white!30]
\tikzstyle{io} = [trapezium, trapezium left angle=70, trapezium right angle=110, minimum width=3cm, minimum height=1cm, text centered, text width = 15cm, draw=black, fill=blue!30]
\tikzstyle{process} = [rectangle, minimum width=3cm, minimum height=0.5cm, text centered, text width=7cm, draw=black, fill=white!30]
\tikzstyle{decision} = [diamond, minimum width=3cm, minimum height=1cm, text centered, draw=black, fill=green!30]
\tikzstyle{arrow} = [thick,->,>=stealth]
 \title{Predicting Molecular Phenotypes with Single Cell RNA Sequencing Data: an Assessment of Unsupervised Machine Learning Models}
    \author{Anastasia Dunca\\West High School, Salt Lake City, UT 84103, e-mail: anastasia.d960@slcstudents.org\\ \vspace{0.1in}\\Project Advisor: Dr. Frederick R. Adler\\Department of Mathematics, University of Utah, UT 84112, e-mail: adler@math.utah.edu}
    \date{August 2021}
\begin{document}
\maketitle
\begin{center}
\color{black} \rule{\linewidth}{0.5mm} 
\end{center}
\begin{abstract}
        According to the National Cancer Institute, there were 9.5 million cancer-related deaths in 2018. A challenge in improving treatment is resistance in genetically unstable cells. The purpose of this study is to evaluate unsupervised machine learning on classifying treatment-resistant phenotypes in heterogeneous tumors through analysis of single cell RNA sequencing (scRNAseq) data with a pipeline and evaluation metrics. scRNAseq quantifies mRNA in cells and characterizes cell phenotypes. One scRNAseq dataset was analyzed (tumor/non-tumor cells of different molecular subtypes and patient identifications). The pipeline consisted of data filtering, dimensionality reduction with Principal Component Analysis, projection with Uniform Manifold Approximation and Projection, clustering with nine approaches (Ward, BIRCH, Gaussian Mixture Model, DBSCAN, Spectral, Affinity Propagation, Agglomerative Clustering, Mean Shift, and K-Means), and evaluation. Seven models divided tumor versus non-tumor cells and molecular subtype while six models classified different patient identification (13 of which were presented in the dataset); K-Means, Ward, and BIRCH often ranked highest with $\sim80\%$ accuracy on the tumor versus non-tumor task and $\sim60\%$ for molecular subtype and patient ID. An optimized classification pipeline using K-Means, Ward, and BIRCH models was evaluated to be most effective for further analysis. In clinical research where there is currently no standard protocol for scRNAseq analysis, clusters generated from this pipeline can be used to understand cancer cell behavior and malignant growth, directly affecting the success of treatment.\\
        
\end{abstract}

\section{Introduction}\label{bg}
\subsection{Tumor Heterogeneity}
Tumor heterogeneity is a state in which cancer cells within the same tumor are functionally different from each other \cite{proposal}. As a patient’s cancer progresses, these cell populations–genetically distinct groups of tumor cells descending from a common ancestor \cite{intra_tumor}–can be identified and their evolutionary trajectory can be followed with response and resistance to treatment. This hypothesis was first made in 1976 with Peter Nowell who suggested that tumors have multiple heterogeneous subclones \cite{proposal}. Individualized cancer treatment was developed in order to stop treating cancer as if it was homogeneous. In particular, tumor heterogeneity can create resistance to cancer treatment. Because the subclonal populations within a tumor can be sensitive or resistant, generalized cancer treatment will become repeatedly less effective with more rounds of treatment. The treatment will only be effective for the portion of the tumor that is not yet resistant. By identifying subclonal populations, targeted therapy may be developed to specifically target the oncogenic causes \cite{targetted_therapy}.
\subsection{Single Cell RNA Sequencing}
Single cell RNA sequencing (scRNAseq) has dominated bioinformatics research with the invaluable information it provides such as complex and rare cell populations, regulatory relationships between genes, and the trajectories of distinct cell lineages in development \cite{sc_rna_tech}. Different cells may express different genes at different levels. These differences are a key into the behavior of cells. The absolute quantity of RNA molecules is very small in a single cell, so experiments often consist of billions of cells. When RNA is extracted from a group of cells, all information about the difference between them is lost, so the data are a mean expression of each gene. To link these results to individuals cells, RNA from each cell is uniquely barcoded and sequenced. For each sequenced molecule, the information attained is the ID of the cell–called the barcode–and it's own unique ID–called the unique molecular identifier. The resulting data from the sequencing is formatted in a genes-by-cells table P, where P[i, j] is an expression of gene i in cell j \cite{single_cell_meets_once_again}.
\subsection{Machine Learning}
Machine learning is a modern computational tool that is at the intersection of linear algebra, statistics, and computer science. Machines automatically learn and improve from experience without being explicitly programmed. In the large umbrella of machine learning, there are many approaches that stem from the two types of learning: supervised and unsupervised learning. Supervised learning is primarily conducted when the data have labels (a target variable on a given set of predictors). Unsupervised learning is primarily conducted when the data lack labels. Given the unlabeled nature of scRNAseq data as well as the lack of independent variables, we here use unsupervised learning.\\
\indent Unsupervised learning is done with cluster analysis and dimensionality reduction. Dimensionality reduction reduces the number of variables (dimensions) in the data while preserving at least 80\% of the variance. Clustering divides the data into groups based on patterns within the data (here, gene expression). In optimal clusters, all data points should be close to each other and the groups should be separated. With scRNAseq data, biologists use clustering to identify how many different cell states there are.\\
\indent In this work, nine different models are evaluated in the five categories of clustering: hierarchical, density based, model based, centroid based, and graph based clustering. The hierarchical clustering methods tested are Agglomerative (average linkage) Clustering, Ward Clustering, and BIRCH Clustering (Balanced Iterative Reducing and Clustering using Hierarchies). The density based clustering methods tested are are DBSCAN (Density Based Clustering of Spatial Applications with Noise). The model based clustering is tested with the Gaussian Mixture Model and Mean Shift. The centroid based clustering is tested with K-Means Clustering. Graph based clustering is tested through Affinity propagation and Spectral clustering.
\subsubsection{Unsupervised Learning Models}
\indent Here, we will present the underlying mathematical methods of each of the unsupervised learning models used in this work (BIRCH, Ward, Spectral, Gaussian Mixture Model, DBSCAN, Affinity Propagation, Agglomerative Clustering, Mean Shift, K-Means). 
\paragraph{BIRCH} The BIRCH model has been proven to be suitable for large databases \cite{birch_cluster}. BIRCH clusters multi-dimensional metric data points to produce the most optimal clusters using dynamic and iterative methods. A primary quality of BIRCH clustering is that it can create a decent structure within one scan of the data, and improve with a few additional scans. It can also handle outlier data points effectively. \\
\indent BIRCH clustering first generates a compact summary of the large dataset that represents as much data as possible \cite{birch_cluster} which is then clustered. BIRCH summarizes larger datasets into smaller, dense regions called Clustering Feature entries. A Clustering Feature Entry is defined as an ordered triple (N, LS, SS). ’N’ represents the number of data points in the cluster, ‘LS’ is the linear sum of the data points and ‘SS’ is the squared sum of the data points in the cluster. A Cluster Feature Tree (CF Tree) is the compact representation that BIRCH creates. A CF Tree is a tree where each leaf node contains a sub-cluster \cite{birch_cluster}.  Every entry in a CF tree contains a pointer to a child node and a CF entry made up of the sum of CF entries in the child nodes. Each leaf node has a maximum number of entries called the threshold. These methods use the Cluster Feature Additivity Theorem \cite{BIRCH_paper}. The parameters for the BIRCH model are the threshold, branching\_factor, and n\_clusters. The threshold is the maximum number of data points a sub-cluster in the leaf node of the CF tree can hold. A branching\_factor specifies the maximum number of CF sub-clusters in each node. n\_clusters is the number of clusters that the dataset should be grouped in once the BIRCH model is complete. \\
The overall process based on the BIRCH model is as follows: \\
 \centerline{\textbf{Given a cluster with N objects described by vectors $\vec{X_i}$:}}\\
 \centerline{\textbf{Define.}}\\
    \begin{equation}
    \begin{aligned}
          \textbf{Centroid: } & \vec{X_0} = \frac{\sum_{i=1}^{N}\vec{X_i}}{N}\\
          \textbf{Radius: }   & R = \Big (\frac{\sum_{i=1}^{N}{(\vec{X_i} - \vec{X_0})^2}}{N}\Big)^{1/2}\\
          \textbf{Diameter: } & D = \Big (\frac{\sum_{i=1}^{N}\sum_{j=1}^{N}(\vec{X_i}-\vec{X_j})^2}{N(N-1)}\Big)^{1/2}\\
    \end{aligned}
    \end{equation}
    \vspace{1cm}\\
    \centerline{\textbf{Given two clusters with N1 and N2 objects: }}
    \centerline{\textbf{Define.}}\\
    \begin{equation}
    \begin{aligned}
        \textbf{Centroid Euclidean Distance: } & D_0 = ((\vec{X_0}-\vec{Y_0})^2)^{1/2}\\
        \textbf{Centroid Manhattan Distance: } & D_1 = |\vec{X_0}-\vec{Y_0}|\\
        \textbf{Average inter-cluster distance: } & D_2 = \Big(\frac{\sum_{i=1}^{N1}\sum_{j=1}^{N2}(\vec{X_i}-\vec{Y_j})^2}{N1N2}\Big)^{1/2}
    \end{aligned}
    \end{equation}
    \centerline{\textbf{Clustering Feature (CF)}}
    \begin{equation}
    \begin{aligned}
        CF = & (N, LS, SS)\\
        N = & |C|, LS = \sum_{i=1}^{N}\vec{X_i}, SS = \sum_{i=1}^{N}\vec{X_i}^2\\
        R = & \Big(\frac{\sum_{i=1}^{N}(\vec{X_i}-\vec{X_0})^2}{N}\Big)^{1/2} = \Big(\frac{\sum_{i=1}^{N}(\vec{X_i}^2-2\vec{X_0}\vec{X_i}+\vec{X_0}^2)}{N}\Big)^{1/2}\\
        = & \Big(\frac{\sum_{i=1}^{N}\vec{X_i}^2-2\sum_{i=1}^{N}\vec{X_i}\vec{X_0}+\sum_{i=1}^{N}\vec{X_0}^2}{N}\Big)^{1/2}\\
         = & \Big(\frac{\sum_{i=1}^{N}\vec{X_i}^2-2\vec{X_0}\sum_{i=1}^{N}\vec{X_i}+{N}\vec{X_0}^2}{N}\Big)^{1/2}\\
        = & \Big(\frac{SS - 2\frac{LS}{N}*LS + N(\frac{LS}{N})^2}{N}\Big)^{1/2}\\
        \end{aligned}
    \end{equation}
    
    \centerline{\textbf{CF Additive Theorem}}
    \begin{equation}
    \begin{aligned}
        C_1: CF_1 = & (N_1, LS_1, SS_1)\\
        C_2: CF_2 = & (N_2, LS_2, SS_2)\\
        CF = &  CF_1 + CF_2\\
           = &  (N_1 + N_2, \vec{LS_1} + \vec{LS_2}, SS_1 + SS_2)\\
    \end{aligned}
    \end{equation}
\indent The first phase of BIRCH clustering consists of inserting nodes into the leaf of the cluster tree in which the algorithm scans data points and traverses down the tree to choose the closest leaf for node \textit{d}. Then, it searches for the closest entry $L_i$ in leaf node. If \textit{d} can be inserted in $L_i$, then the Cluster Feature is updated with vector $L_i$. If \textit{d} cannot be inserted to $L_i$, but there is space to insert a new entry, then a new entry will be inserted. If neither of the previous conditions are true, then the node will split. Once inserted, the nodes along the path to the root are updated; if there is splitting instead, a new entry needs to be inserted into the parent node. \\
\indent The next phase of BIRCH is global clustering which uses existing clustering algorithms on sub-clusters at leaf nodes which may treat each sub-cluster as a point and perform clustering on these points. Clusters are produced closer to data distribution patterns. The last phase is cluster refinement. The algorithm redistributes (re-labels) all data points and clusters produced by global clustering.
\paragraph{Ward} The Ward model looks at cluster analysis as an analysis of variance problem \cite{ward_penn}. Ward's approach is agglomerative in which it begins at the smallest individual structure and then progresses to a global structure. It starts out with \textit{n} clusters of size 1 and continues until all observations are in one cluster. \\
Ward’s method defines the distance between cluster A and cluster B as:
    \begin{equation}
        \Delta(A, B) = \sum\limits_{i\in{A\bigcup{B}}}\norm{\vec{X_i}-\vec{m_{A\bigcup{B}}}}^2-\sum\limits_{i\in{A}}\norm{\vec{X_i}-\vec{m_{A}}}^2 - \sum\limits_{i\in{B}}\norm{\vec{X_i}-\vec{m_{B}}}^2
    =\frac{n_An_B}{n_A+n_B}\norm{\vec{m_A}-\vec{m_B}}^2
    \end{equation}
    where $\vec{m_j}$ is the center of cluster j, and $n_j$ is the number of points in it. Function $\Delta(A, B)$ is called the \textbf{merging cost} of combining clusters \textit{A} and \textit{B}. \\
\indent With general hierarchical clustering, the sum of squares starts out at zero and then grows as it merges clusters. The method keeps the growth as small as possible \cite{ward_method}. Given two pairs of clusters whose centers are equally far apart, Ward’s method will prefer to merge the smaller clusters. Let $X_{ijk}$ denote the value for variable k in observation  j belonging to cluster i. The Ward model will sum over all variables, and all of the units within each cluster. Then each individual observation is compared for each variable against the cluster means for that variable.\\
    \centerline{Here we review some mathematical definitions:} 
    \begin{equation}
    \begin{aligned}
    \text{Error Sum of Squares: } ESS & = \sum_i \sum_j \sum_k |X_{ijk} - \overline{x}_{i.k}|^2\\
    \text{Total Sum of Squares: } TSS & = \sum_i \sum_j \sum_k  |X_{ijk} - \overline{x}_{..k}|^2\\
    \text{R-Square }: r^2 & = \frac{TSS-ESS}{TSS}
    \end{aligned}
    \end{equation}
\indent Using Ward's method, $n-1$ clusters are formed, one of size two and the remaining of size 1. The error sum of squares and $r^2$ values are computed. The pair of data points that have the smallest error sum of squares or largest $r^2$ value will compose the first cluster. The second step of Ward's method consists of forming $n-2$ clusters from $n-1$ clusters defined in the previous step. The value of $r^2$ is maximized again to determine which data points form the next cluster. Iteratively, clusters are formed on the basis of keeping the $r^2$ value at a max or error sum of squares at a minimum. The algorithm stops once each sample unit is combined into a single large cluster of size n, or the specified number of clusters are formed. 
\paragraph{Spectral}  clustering focuses on clustering points that are connected rather than compact. Points that are next to each other are put into the same cluster. While there may be points that have a smaller distance between them, if they are not connected, the spectral clustering algorithm will not cluster them together \cite{spectral_foundation}. The model treats the data points as nodes of a graph. The nodes are mapped to a low dimensional space that can be easily partitioned to form multiple clusters. \\
\indent Given a set of data points $x_1,…x_n$ and a notion of similarity $s_{ij} >= 0$ between all pairs of data points $x_i$ and $x_j$, the goal of Spectral clustering is to divide the data points in several groups in which the points in the same group are similar and points in different groups are dissimilar to each other. Without extra information on the similarity of each data point, a similarity graph takes the form of $G = (V, E)$. Each vertex $v_i$ in the graph represents a data point $x_i$. Two vertices are connected if the similarity $s_{ij}$ between two data points is greater than a certain threshold, and the edge is weighted by $s_{ij}$. After constructing the graph, the goal for clustering can be rephrased: find a partition of the graph such that the edges between different groups have very low weights and the edges within a group have high weights. \\
\indent Spectral clustering first computes a similarity graph. Similarity graphs can be computed with an e-neighborhood graph in which a distance threshold between each point is used to determine if points are connected; k nearest neighbors graph in which a point is connected to another if it is the closest neighbor to that point; or a fully connected graph in which all points are connected with each other and edges are weighted by the similarity $s_{ij}$ which models local neighborhood relationships \cite{spectral_overview}. Then, the data points are projected onto a low-dimensional space using Graph Laplacians. The graph Laplacians find eigenvalues and eigenvectors in order to embed data points into low-dimensional space. Lastly, clusters are formed by using the eigenvector corresponding to the 2nd eigenvalue to assign values to each node. To get two distinct clusters, nodes higher than a specific value are assigned to one cluster whereas the others nodes are assigned to another one. 
\paragraph{Gaussian Mixture Model}is a probabilistic model for representing normally distributed subpopulations within an overall population. Mixture models do not require knowledge of the subpopulation's data points, but rather allows the model to learn the subpopulations automatically. The most commonly used distribution in modeling real world unimodal data is the Gaussian distribution. Modeling multimodal data as a mixture of unimodal Gaussian distributions is intuitive. Therefore, Gaussian Mixture Models maintain many benefits of Gaussian models, which contributes to their practicality and efficiency when modeling very large datasets \cite{gmm}.\\
\indent A Gaussian Mixture is a function comprised of several Gaussians–each identified by $k\in{{1,...,K}}$ where K is the number of desired clusters. Each Gaussian k in the mixture possesses the following parameters: a mean $\mu$ that defines the centroid; a covariance $\Sigma$ that defines its width; a mixing probability $\pi$ that defines the size of the Gaussian function. Each Gaussian explains the data contained in each of the three clusters available \cite{gmm2}. The mixing coefficients, $\pi_k$ are probabilities that must meet this condition: $\sum\limits_{k=1}^K\pi_k=1$. Each Gaussian must also fit the data points belonging to each cluster. The Gaussian density function is: 
\begin{equation}
    N(x|\mu, \Sigma)=\frac{1}{(2\pi)^{D/2}|\Sigma|^{1/2}}exp(-\frac{1}{2}(x-\mu)^T\Sigma^{-1}(x-\mu))
\end{equation}
\indent Here, the variable x represents the data points; D is the number of dimensions of each data point. $\mu$ and $\Sigma$ are the mean and covariance. Differentiating this equation with respect to the mean and covariance, then equating it to zero, will reveal the optimal values for those parameters. This learning technique for Gaussian Mixture Models is also called Expectation-Maximization. In general, Expectation-Maximization tries to use existing data to determine the optimum values for these variables and then finds the model parameters. The Gaussian Mixture Model as two steps: E-step: the available data is used to estimate the values of the missing variables; M-step: the complete data is used to update the parameters. For E-Step, at each point $x_i$, the probability that it belongs to cluster $C_1, C_2, ..., C_k$ is calculated: 
\begin{equation}
    r_iC = \frac{\text{Probability } x_i \text{ belongs to C }}{\text{ Sum of Probability } x_i \text{ belongs to } C_1, C_2, ..., C_k}=\frac{\pi_CN(x_i;\mu_C, \Sigma_C)}{\Sigma_C, \pi_C, N(x_i; \mu_C, \Sigma_C)}
\end{equation}
The M-Step goes back and updates the $\pi, \mu \text{ and } \Sigma$ values:
\begin{equation}
    \pi = \frac{\text{Number of points assigned to cluster}}{\text{Total number of points}}
\end{equation}
\indent Then the mean and covariance matrix are updated based on values assigned to the distribution in proportion with probability values for the data point. Based on the updated values generated from this step, the new probabilities for each data point are calculated and updated iteratively \cite{better_gmm}.
\paragraph{DBSCAN} is Density-Based Spatial Clustering of Applications with Noise. The algorithm focuses on point density in which points close in distance are grouped together whereas outliers are found in regions of low point density \cite{dbscan_theory}. 
For any point p, the neighboring points with a radius $\epsilon$ are part of the $\epsilon$ neighborhood: 
\begin{equation}
N_{\epsilon}(p):{q|d(p,q)<= \epsilon}
\end{equation}
 \indent The $\epsilon$-neighborhood for a point $p$ is the set of all q such that the distance between $p$ and $q$ is less than or equal to $\epsilon$. Points are classified as either high or low density depending on the value of MinPts. If the $\epsilon$-neighborhood contains at least MinPts then the point is considered high-density. Otherwise, it’s low density. Given a dataset and values for $\epsilon$ and MinPts, DBSCAN can categorize the points into three exclusive groups: core points, border points, and noise points. A core point has at least MinPts within the $\epsilon$ neighborhood; a border point has fewer than MinPoits but is in the $\epsilon$ neighborhood of a core point. A noise point fails to satisfy the two previous conditions. A point $q$ is directly density-reachable from another point $p$ if $p$ is a core point and $q$ is in $p$’s $\epsilon$-neighborhood. A point $q$ is indirectly density-reachable from another point $q$ if there exists a chain of points $q_1$…$q_n$, $q_1=p$, $q_n = q$ such that $q_{n+1}$  is directly density-readable from $q_1$. A point $q$ is density-connected to a point $p$ if there is a point $o$ such that both $p$ and $q$ are density-reachable from $o$. \\
\indent Let $X = {x_1, x_2, x_3…x_n}$ be a data set. DBSCAN starts with an arbitrary starting point. It extracts the neighborhood of this point using the value of $\epsilon$ where all points are under distance $\epsilon$ from the neighborhood. If the neighborhood achieves the MinPts threshold then the clustering process starts and the point is marked as visited; if it does not reach that threshold, then the point is labeled as noise. If the point is part of a cluster, then its $\epsilon$ neighborhood is also the part of the cluster and the above procedure is repeated for all $\epsilon$ points. This is repeated until all points in the cluster are determined. All density-reachable points are included in clusters and are not labeled as noise. A new unvisited point is retrieved and processed–classified as either another cluster or noise. This process continues until all points have been visited \cite{dbscan_math}.  
\paragraph{Affinity Propagation} is a graph based clustering algorithm that finds “exemplars” or members of the input dataset that are representative of clusters. It takes a similarity matrix as input and identifies exemplars based on criteria. Messages are exchanged between the data points until a high-quality set of exemplars are obtained. With the intake of a dataset, affinity propagation calculates a similarity matrix, responsibility matrix, availability matrix, and criterion matrix. A similarity matrix is calculated with: 
\begin{equation}
s(i, j) = -||x_i - x_j||^2
\end{equation}\\
which is the negative euclidean distance between the two instances. The greater the distance, the smaller similarity. For an element $i$, affinity propagation looks for another element $j$ for which $s(i, j)$ is the highest. \\
\indent A responsibility matrix $r(i, k)$ quantifies how well-suited element k is, to be an exemplar for element $i$, taking into account the nearest contender k’ to be an exemplar for $i$. 
\begin{equation}
r(i, k) \leftarrow s(i, k) - max_{{k’ \neq k}}(a(i, k’) + s(i, k’))
\end{equation}
\indent The R matrix is initialized with zeros. It represents the relative similarity between $i$ and $k$ and quantifies how similar $i$ is to $k$ compared to some $k’$ taking into account the availability of $k’$. The responsibility of $k$ towards $i$ will decrease as the availability of some other $k’$ other to $i$ increases.\\
\indent An availability a(i, k) quantifies how appropriate it is for $i$ to choose $k$ as its exemplar, valuing the support from other elements that $k$ should be an exemplar. 
\begin{equation}
a(i, k) \leftarrow min ( 0, r(k, k) + \sum_{i’ \notin \{i, k\}} max(o, r(i’, k))) \text{ for } i \neq k
\end{equation}
\indent Availability is self-responsibility of k and the positive responsibilities of k towards elements other than i. If self responsibility is negative, that means k is more suitable to belong to another exemplar rather than being an exemplar. The maximum value of an availability is 0. 
Self availabilities are: 
\begin{equation}
a(k, k) \leftarrow \sum_{i’ \neq k} max(0, r(i’, k))
\end{equation}
a(k, k) reflects accumulated evidence that point k is suitable to be an exemplar based on the positive responsibilities of k towards other elements.
\indent Responsibility and Availability Matrices are iteratively updated for a fixed number of iterations after changes in the values obtained fall below a threshold. \\
\indent Lastly, there is a criterion matrix that is calculated after the updating is terminated. Criterion matrix C is the sum of R and A. An element i will be assigned to an exemplar k which is not only highly responsible but also highly available to i. 
\begin{equation}
c(i, k) \leftarrow r(i, k) + a(i, k)
\end{equation}
\indent The element with the highest criterion value in each row of the dataset would be an exemplar. Elements corresponding to the rows which share the same exemplar are clustered together \cite{aff_prop_math}. 
\paragraph{Agglomerative Clustering Average Linkage} assigns each observation to its own cluster. As opposed to Ward's method which uses a "similarity" and/or "variance" criterion instead of a distance metric \cite{agglomerative_hierarchical}, average linkage is defined to be the average distance between data points in the first cluster and data points in the second cluster. On the basis of this definition of distance between clusters, at each stage of the process it combines the two clusters that have the smallest average linkage distance. These steps are iterated until there is one cluster left. Before clustering, a proximity matrix containing the distance between each point using a distance point is calculated. Then, the matrix is updated to display the distance between each cluster. In this research, average linkage is used to measure the distance between each cluster. \cite{agglom}\\
\indent If the data points are represented as normalized vectors in Euclidean space, the cohesion $G$ of a cluster $C$ is defined as the average dot product: 
\begin{equation}
\begin{aligned}
& G(C) = \frac{1}{n(n-1)}(\gamma(C)-n) \\ 
& \text{where}\\
& n = |C|, \\
& \gamma(C) = \sum_{v \in C} \sum_{w \in C} <v, w> \\
& \text{and} <v, w> \text{is the dot product of v and w}
\end{aligned}
\end{equation}
If an array of cluster centroids $p(C)$, where $p(C) = \sum v \in C v$ for the currently active clusters, then $\gamma$ of the merger of $C_1$ and $C_2$ can be computed recursively: 
\begin{equation}
\begin{aligned}
\gamma(C_1 + C_2)  = & \sum_{v \in C_1 + C_2} \sum_{w \in C_1 + C_2}<v, w>\\
 = & <\sum_{v \in C_1 + C_2}v, \sum_{w \in C_1 + C_2}w>\\
 = & <p(C_1 + C_2), p(C_1 + C_2)>\\
\end{aligned}
\end{equation}
\indent Overall, the model first computes all $n^2$ similarities for the singleton clusters, and sort them for each cluster. In each merge iteration, a pair of clusters with the highest cohesion is identified and merged; then cluster centroids, $\gamma$ s, and cohesions of the possible mergers of the just created cluster with the remaining clusters are updated. For each cluster, the sorted list of merge candidates are also updated by deleting the two just merged clusters and inserting its cohesion with the just created cluster. The iterations continue until all points have been merged into a single remaining cluster. \cite{stanford_avg_link}
\paragraph{Mean Shift} is based on the kernel density estimation which estimates the underlying distribution of a dataset. It works by placing a kernel–or weighting function–on each point in the dataset. Adding all of the kernels up generate a probability surface or density function. The mean shift model exploits the kernel density estimation by predicting the behavior of the points if they were all at the peak on the KDE surface by shifting each point “uphill” until it reaches a peak. It starts by making a copy of the original dataset and freezes the original points. The copied points are shifted against the original points. Each point is moved closer to the nearest KDE surface peak in each iteration. \cite{spin_ms}\\
\indent Given $n$ data points $x_i, i = 1, …, n$ on a d-dimensional space $R^d$, the multivariate kernel density estimate obtained with kernel $K(x)$ and window radius $h$ is: 
\begin{equation}
f(x) = \frac{1}{nh^d}\sum_{i=1}^n K\Big(\frac{x-x_i}{h}\Big)
\end{equation}
The gradient of the density estimator is:
\begin{equation}
\begin{aligned}
\nabla f(x) & = \frac{2c_{k,d}}{nh^{d+2}} \sum_{i=1}^n(x_i-x)g\Big(\Big|\Big|\frac{x-x_i}{h}\Big|\Big|^2\Big)\\
& = \frac{2c_{k,d}}{nh^{d+2}} \Big[\sum_{i=1}^n(x_i-x)g\Big(\Big|\Big|\frac{x-x_i}{h}\Big|\Big|^2\Big)\Big] \Big[\frac{\sum_{i=1}^n x_ig(||\frac{x-x_i}{h}||^2)}{\sum_{i=1}^n g(||\frac{x-x_i}{h}||^2)}-x\Big]
\end{aligned}
\end{equation}
where $g(s) = -k’(s)$. The first term is proportional to the density estimate at x computed with kernel $G(x) = c_{g, d}g(||x||^2)$ and the second term: 
\begin{equation}
m_h(x) = \frac{\sum_{i=1}^n x_ig(||\frac{x-x_i}{h}||^2)}{\sum_{i=1}^n g(||\frac{x-x_i}{h}||^2}-x
\end{equation} 
 is the \textit{mean shift}. The mean shift vector points toward the direction of the maximum increase in the density. The mean shift procedure, obtained by: 
\begin{itemize}
\item computation of the mean shift vector $m_h(x^t)$
\item translation of the window $x^{t+1} = x^t + m_h(x^t)$
\end{itemize}
is guaranteed to converge to a point where the gradient of density function is zero–or at a peak. \cite{ms_eq}\\
\indent With each iteration, the data points will move closer to where the most points are, which will lead to the cluster center. The algorithm stops when each point is assigned to a cluster \cite{gg}.
\paragraph{K-Means} finds a user-specified number of clusters (k) which are represented by their centroids. It starts out by choosing k initial centroids. Each point is then assigned to the closest centroid and each collection of points assigned to a centroid is a cluster. The centroid of each cluster is then updated based on the points assigned to the cluster. The assignment and update steps are iterated until no points change clusters or the centroids remain the same. \\
\indent In K-Means proximity measures are used to define the similarity/dissimilarity between two data points. Similarity measure is large if the features are similar whereas the dissimilarity measure is small if features are similar. The distance is measured in Euclidean space\cite{k_medium}: 
\begin{equation}
\begin{aligned}
d(p, q) = \sqrt{\sum_{i=1}^n(q_i - p_i)^2}
\end{aligned}
\end{equation}
\indent To measure the quality of clustering, the sum of the squared error (SSE) is used. The error of each data point is calculated which is also its Euclidean distance to the closest centroid, and then the total sum of the squared errors is computed. A smaller squared error means the centroid is a satisfactory representation of the points within the cluster\cite{k_understanding}: 
\begin{equation}
J = \sum_{j=1}^k \sum_{i=1}^n || x_i^{(j)} - c_j || ^ 2
\end{equation}
\subsection{Comparison of Models Table}
   \begin{table}[H]
    \centering
    \begin{tabular}{c|c|c}
       Category  &  Model(s) & Key Differences Between Models\\
       \hline \hline
        Hierarchical  & BIRCH, Ward, Agglomerative & Metrics for similarity\\
        \hline
        Graph Based & Spectral, Affinity Propagation & Metrics for distance and connectivity\\
        \hline 
        Density Based & DBSCAN \\
        \hline 
        Model Based & Mean Shift, Gaussian Mixture Model & Metrics for probability\\
        \hline
        Centroid Based & K-Means
    \end{tabular}
    \caption{Models Evaluated}
\end{table}

\subsection{Contributions of this Work}
\indent Single cell RNA sequencing is an emerging technology that requires modern, efficient methods of data analysis to properly handle big data and produce useful results for computational research. Currently, there is no standard scRNAseq analysis pipeline, nor has any thorough study been done on which clustering models respond best to scRNAseq data. In this study, we investigate which unsupervised learning methods are most effective and efficient for scRNAseq data analysis through rigorous evaluation. After experimentation, it was seen that for multiple tasks the highest performing models were BIRCH, K-Means, and Ward. With these models, an optimized pipeline was constructed that can be used for future data analysis. With knowledge of the highest performing models, it is also possible to combine the underlying methodology to create an even more effective model that can respond best to scRNAseq data. \\
\indent The paper is organized as follows. In Section 2, we discuss our methodology for analysis and evaluation including brief backgrounds for each computational tool and technique. In Section 3, our findings are presented with visualizations of clustering accuracy, analysis over 100 simulations, and a brief discussion of randomization tests conducted to validate the evaluation method. Section 4 discusses the conclusions that can be drawn from our work. 
\section{Methodology}\label{methods}
In this section, we present the materials and methods used to analyze the scRNAseq. We will also discuss the rationale behind the methods along with more mathematical principles in the tools used.  
\subsection{Pipeline}
\indent The goal for analysis in this work is to separate the entire population of the dataset into specific subpopulations: tumor v. nonTumor, patient ID, and molecular subtype. Each predicted visualization created by the clustering models is evaluated on accuracy by comparing true state to predicted state. In our study, 100 simulations of the clustering models and evaluation are run, then the average accuracies for each model are ranked. Then, the actual true states are projected onto the cluster to examine the true state of each data point and the general structure of the clusters (Figures 6, 7, 8).
    \begin{figure}[H]
        \centering
        \begin{tikzpicture}[node distance=0.8cm]
            \node (start) [startstop] {\tiny SCRNASEQ ANALYSIS AND EVALUATION PIPELINE};
            \node (data) [process, below of=start] {\tiny DATA WRANGLING};
            \node (dimensionality) [process, below of=data] {\tiny DIMENSIONALITY REDUCTION};
            \node (dataviz) [process, below of=dimensionality] {\tiny PRELIMINARY DATA VISUALIZATION};
            \node (clustering) [process, below of=dataviz] {\tiny CLUSTERING};
            \node (comparison) [process, below of=clustering] {\tiny COMPARISON TO TRUE STATE};
            \node (analysis) [startstop, below of=comparison] {\tiny ANALYSIS OF ACCURACY};
            \node (numbers) [process, below of=analysis]{\tiny OBSERVE \# OF POPULATIONS SEPARATED};
            \node (mapOne) [process, below of=numbers, yshift=-0.2cm] {\tiny MAPPING COORDS TO CELL BARCODES};
            \node (mapTwo) [process, below of=mapOne] {\tiny MAPPING COORDS TO CLUSTER LABEL};
            \node (bothMap) [process, below of=mapTwo] {\tiny MAPPING LABELS TO CELL BARCODES};
            \node (files) [process, below of=bothMap] {\tiny CREATE FILES FOR PREDICTED CELLS};
            \node (compare) [process, below of=files] {\tiny COMPARE TO TRUE CELLS FOR EACH POPULATION};
         
            \node (simulations) [process, below of=compare] {\tiny 100 RUNS};
            \node (results) [startstop, below of=simulations] {\tiny ACCURACY AND STRUCTURE EVALUATION};
            
            \draw [arrow] (start) -- (data);
            \draw [arrow] (data) -- (dimensionality);
            \draw [arrow] (dimensionality) -- (dataviz);
            \draw [arrow] (dataviz) -- (clustering);
            \draw [arrow] (clustering) -- (comparison);
            \draw [arrow] (comparison) -- (analysis);

            \draw [arrow] (analysis) -- (numbers);
            \draw [arrow] (numbers) -- (mapOne);
            \draw [arrow] (mapOne) -- (mapTwo);
            \draw [arrow] (mapTwo) -- (bothMap);
            \draw [arrow] (bothMap) -- (files);
            \draw [arrow] (files) -- (compare);
            \draw [arrow] (compare) -- (simulations);
            \draw [arrow] (simulations) -- (results);
        \end{tikzpicture}
        \caption{Observation Flowchart}
        \label{processflow}
  \end{figure}
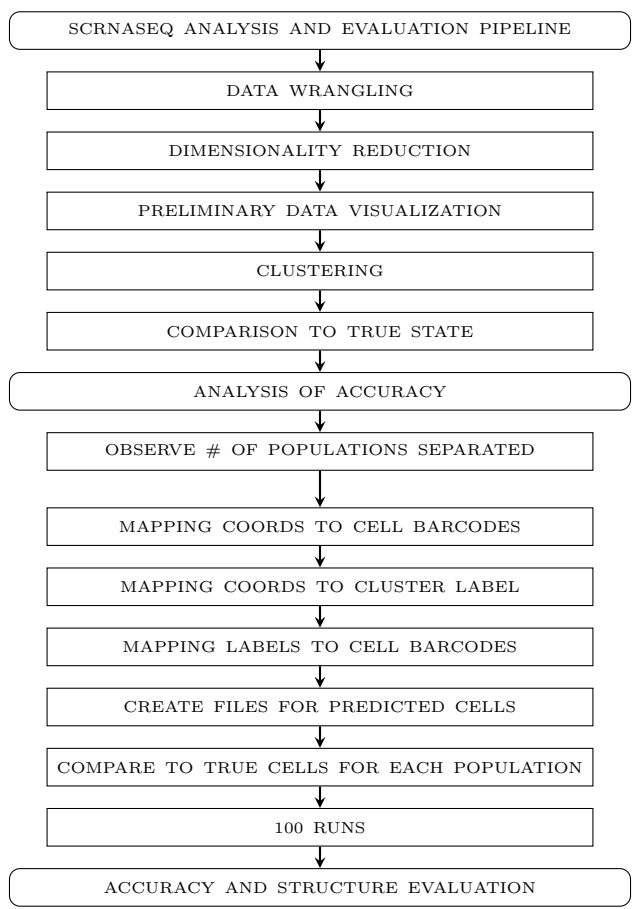
    
\subsection{Data}
The data used in this study is a gene expression matrix with gene counts per million cells. The initial shape of the dataset was 57822 genes by 517 cells; after filtering outliers, the shape analyzed was 6563 genes by 517 cells. Three goals were assigned: separation by cell state (tumor versus nonTumor), molecular subtype (estrogen receptor positive, double positive, triple negative breast cancer, and human epidermal growth factor), and patient identification (BC01, BC02, BC03, BC03\_LN, BC04, BC05, BC06, BC07, BC07\_LN, BC08, BC09, BC10, BC11). These data were sequenced from the cell lines of estrogen receptor positive and negative cells.
\subsection{Materials}
\indent The computational tools primarily used in this research were Python 3, JupyterNotebooks, and Anaconda. The libraries used were sklearn, numpy, PCA, matplot, umap, pandas, and scipy. 
\subsection{Exploratory Data Analysis and Data Filtering}
\indent The necessary libraries and machine learning tools need to be imported before any algorithms can be made. The scRNAseq dataset is loaded as well as the features (genes) and barcodes (cells) labels are assigned. For filtering and EDA (exploratory data analysis), it is important to understand the dimensions of the data as well as have a good image of the dataset. Plotting the current unfiltered data will allow for filtering opportunities and observations of total noise. In this case, UMI’s (Unique Molecular Identifiers) per cell and per gene will be identified. This will be used to choose high and low express genes to filter out. The purpose of identifying and eliminating the outliers is so that as the machine learning algorithms learn, the outlier values will not have a heavy impact of the decisions that the model makes. With technical errors such as false drop out rates, it is important to clean the data for optimal performance of the models.\\
\indent Based on the EDA, outliers are genes with an expression of below 100 and above 1,000,000. Only extreme outliers should be filtered out. Then, the coefficient of variation is plotted to define the highly variable genes in the dataset. Filtration is performed twice more by keeping only the highly variable genes and removing cells with very low expression. 
\subsection{Dimensionality Reduction with Principal Component Analysis}
A Scree Plot is a diagnostic tool to check whether PCA works well on the current data. Principal components are made in the order of the amount of variation they cover: the first component captures the most and it decreases from there \cite{onestop_pca}. In PCA, there are as many principal components as features. The scree plot should be an indicator of what principal components to keep; ideally, the graph should somewhat look like an elbow, and the start of the elbow is the cut off (or the maximum number of principal components to include) \cite{interpret_pca}. The y axis is eigenvalues which means the amount of variation kept with each component. The kept principal components should be able to represent at least 80\% of the total variance in the dataset.\\

\indent The total number of features in the dataset is approximately 600, so naturally our scree plot is very zoomed out. However, minor skews can be observed around 50-100 components. \\
\indent PCA is a dimensionality reduction technique which performs feature extraction on a dataset. Feature extraction creates ``new'' independent variables from selected independent variables that combine each of the old variables and order them by how well they predict the dependent variable. From there, dimensions are reduced based on which important new independent variables are kept \cite{onestop_pca}. In the end, the most valuable parts of all previous variables are kept when the most important new independent variables are selected.\\
\begin{wrapfigure}[12]{r}{0.6\textwidth}
    \centering
    \includegraphics[width=0.4\textwidth]{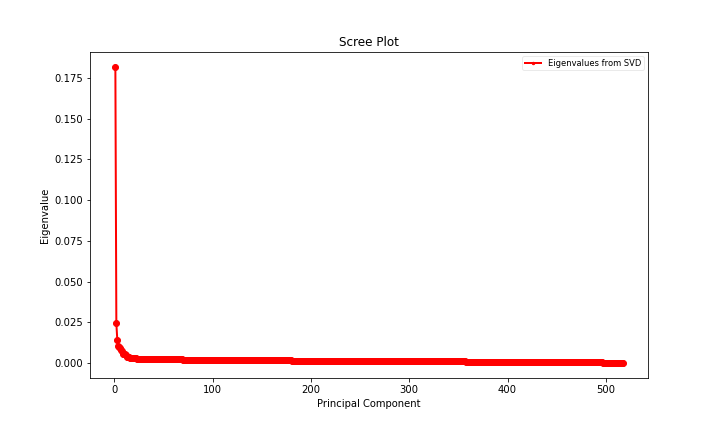}
    \caption{Scree Plot}
    \label{fig:2scree}
\end{wrapfigure}
\indent PCA works by calculating a matrix that summarizes how each of the variables relate to one another. This matrix breaks into two components: direction and magnitude. Then, the data are aligned with the important directions. With direction selection then data compression, the dimensionality of the feature space is reduced; however, the data are only transformed into different directions, so all original variables are still in the model. PCA is preprocessing step that keeps most variance in the data with little information loss, but greatly speeds up time efficiency in later steps such as UMAP and clustering. In this step, we kept 100 principal components.
\subsection{Preliminary Visualization with UMAP}
Now that the data has gone through feature extraction and the most variable components are kept, the data can be projected with UMAP through the creation of a low dimensional embedding. UMAP stands for Uniform Manifold Approximation and Projection for Dimension Reduction. It provides a general framework for approaching manifold learning and dimension reduction. It is mainly used for topological data analysis.\\
\indent The UMAP algorithm works takes all the data points {$x_1$ ,....,$x_n$} and then its k nearest neighbors\cite{penn_umap}. Let $L_i$ be the diameter of the neighborhood of $x_i$ and let $p_i$ be the distance from $x_i$ to its nearest neighbor. Then, a weighted graph is formed by taking $x_j$ in the k-nearest neighborhood of $x_i$ and defining $w_i$($x_i$, $x_j$) = $\frac{exp(-(d(x_i, x_j)-p_i)}{L_i}$. To symmetrize it, $w(x_i, x_j)= w_i(x_i, x_j) + w_j(x_i, x_j) - w_i(x_i, x_j)w_j(x_i, x_j)$ is used. Then, given two weights w and w' on the dataset, cross entropy is represented by:
\begin{equation}
C(w, w') = \sum\limits_{i~j} w(i,j)log\Big(\frac{w(i, j)}{w'(i, j)}\Big) + (1 - w(i, j)) log\Big(\frac{1-w(i, j)}{1-w'(i, j)}\Big)
\end{equation}
w represents the weights computed from the dataset and $w'$ represents the weights computed from our low-dimensional embedding. The cross entropy is the sum of uncoupled terms, one for each edge, so stochastic gradient descent is applied by choosing one term to approximate the gradient at each step. The weights $w'(i, j)$ depend only on the points within the neighborhoods of $x_i$ and $x_j$ which is $2k-1$ points \cite{penn_umap}. The weights themselves represent how likely two data points are connected. To determine connectedness, UMAP extends a radius outwards from each point, connecting points when those radii overlap. UMAP chooses a radius locally, based on the distance to each point's nth nearest neighbor. UMAP then makes the graph "fuzzy" by decreasing the likelihood of connection as the radius grows. Finally, by assuming each point must be connected to its closest neighbor, UMAP ensures that both local and global structure is preserved \cite{umap_overview}. \\
\begin{wrapfigure}{r}{0.6\textwidth}
    \centering
    \includegraphics[width = 0.35\textwidth]{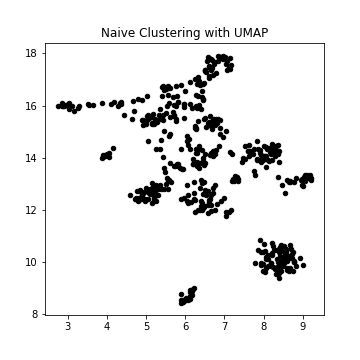}
    \caption{Preliminary Clustering}
\end{wrapfigure}
\indent UMAP simply constructs a high dimensional graph representation of the data then optimizes a low-dimensional graph to be as structurally similar as possible \cite{umap_overview}. The underlying mathematical concepts described above are how UMAP makes its high dimensional graph. Then, UMAP makes the graph "fuzzy" by decreasing the likelihood of connection as the radius grows. By stipulation that each point must be connected to at least its closest neighbor, UMAP preserves local structure in balance with global structure.
\subsection{Parameterization}
\indent After principal components have been selected with PCA and dimensions have been reduced with UMAP, multiple clustering algorithms can be tested for data analysis and for a consistency evaluation. If the partitions of the clusters are relatively consistent amongst the models, this indicates that the parameters used are optimal. 9 clustering algorithms will be tested: Affinity Propagation, Mean Shift, Spectral Clustering, Ward, Agglomerative Clustering, DBSCAN, Birch, K-Means, and the Gaussian Mixture Model.\\
\indent The first step is to define all of the parameters into a dictionary such as quantile (where a sample is divided into adjacent subgroups), epochs (one complete presentation of the dataset to be learned to a learning machine), damping (measure to describe how rapidly oscillations decay from one bounce to the next), preference (task of learning to predict an order relation on collection), n\_neighbors (how many points are assumed to be connected to one point), and n\_clusters (the amount of clusters differentiated). These parameters are optimized with fine tuning and iterative examination.
\subsection{Comparison}
Below are three diagrams that show the cluster analysis of the nine models. Each different color shows a different cluster. From these, we can discuss an aesthetic analysis of how each model performed based on the look of the clusters themselves. 
\begin{figure}[H]
\begin{subfigure}{0.35\textwidth}
    \centering
    \includegraphics[width=\textwidth]{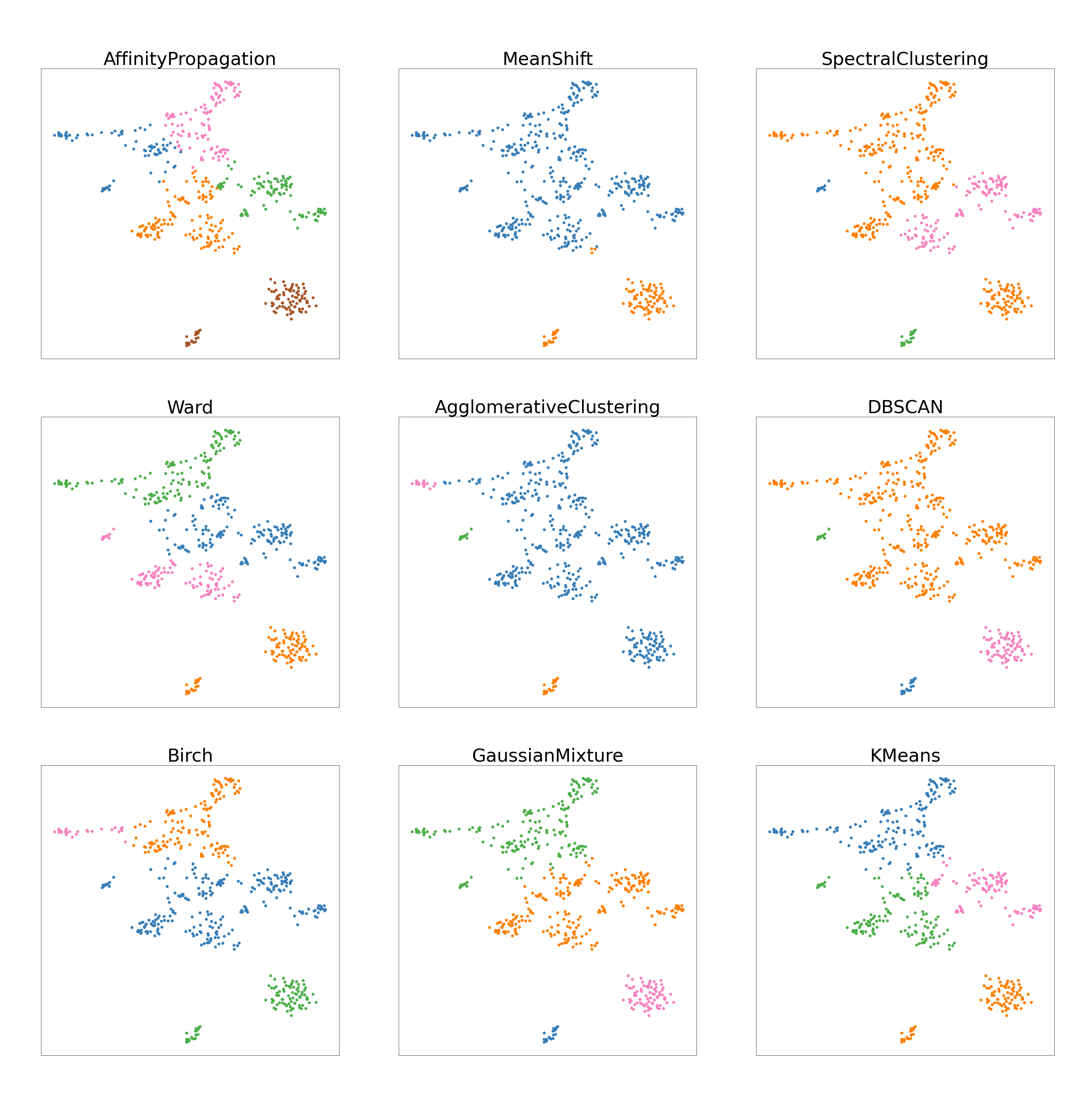}
    \caption{Molecular Subtype}
    \label{fig:acmol}
\end{subfigure}%
\begin{subfigure}{0.35\textwidth}
    \centering
    \includegraphics[width=\textwidth]{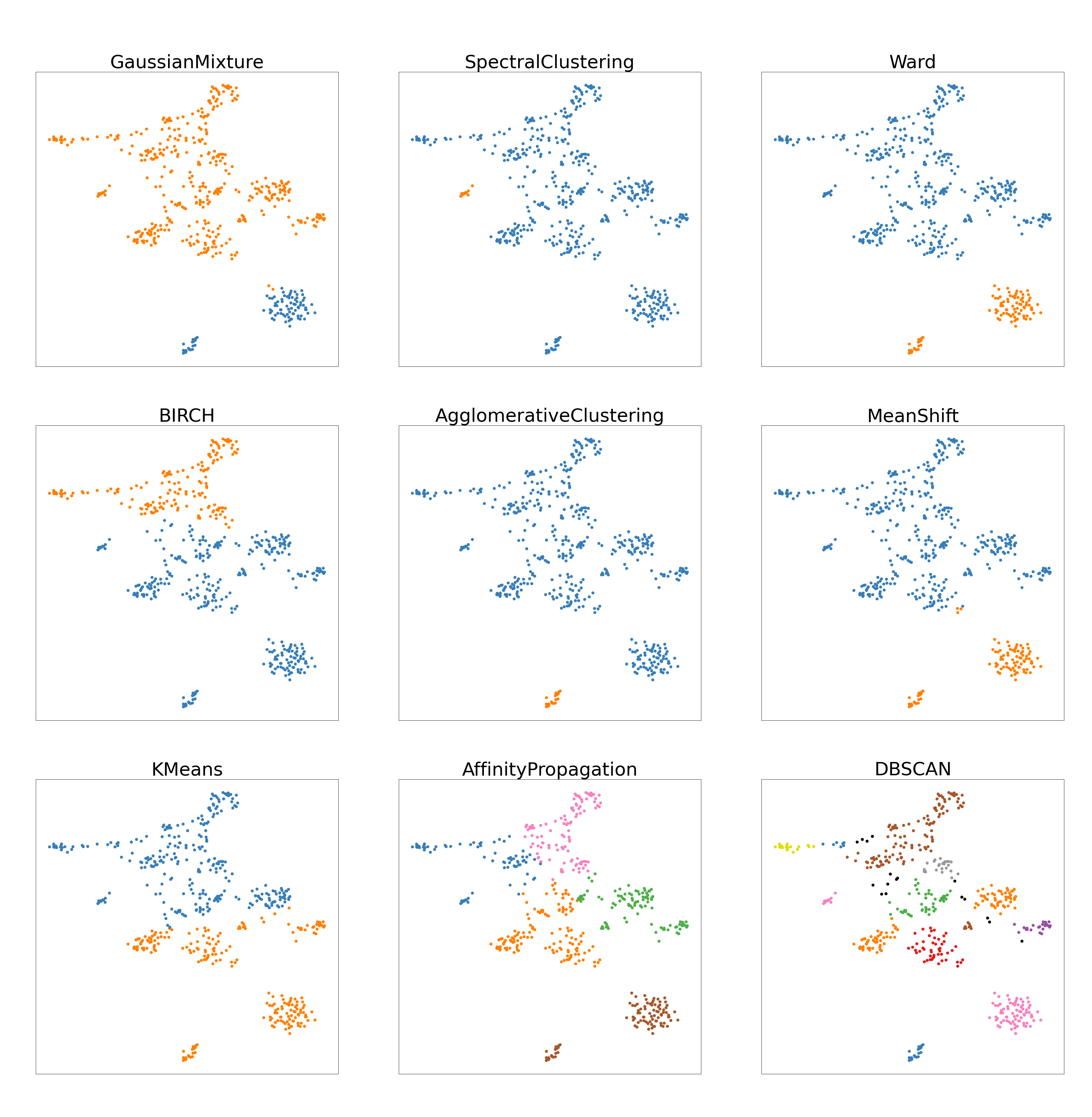}
    \caption{Tumor Cell}
    \label{fig:actumor}
\end{subfigure}%
\begin{subfigure}{0.35\textwidth}
    \centering
    \includegraphics[width=\textwidth]{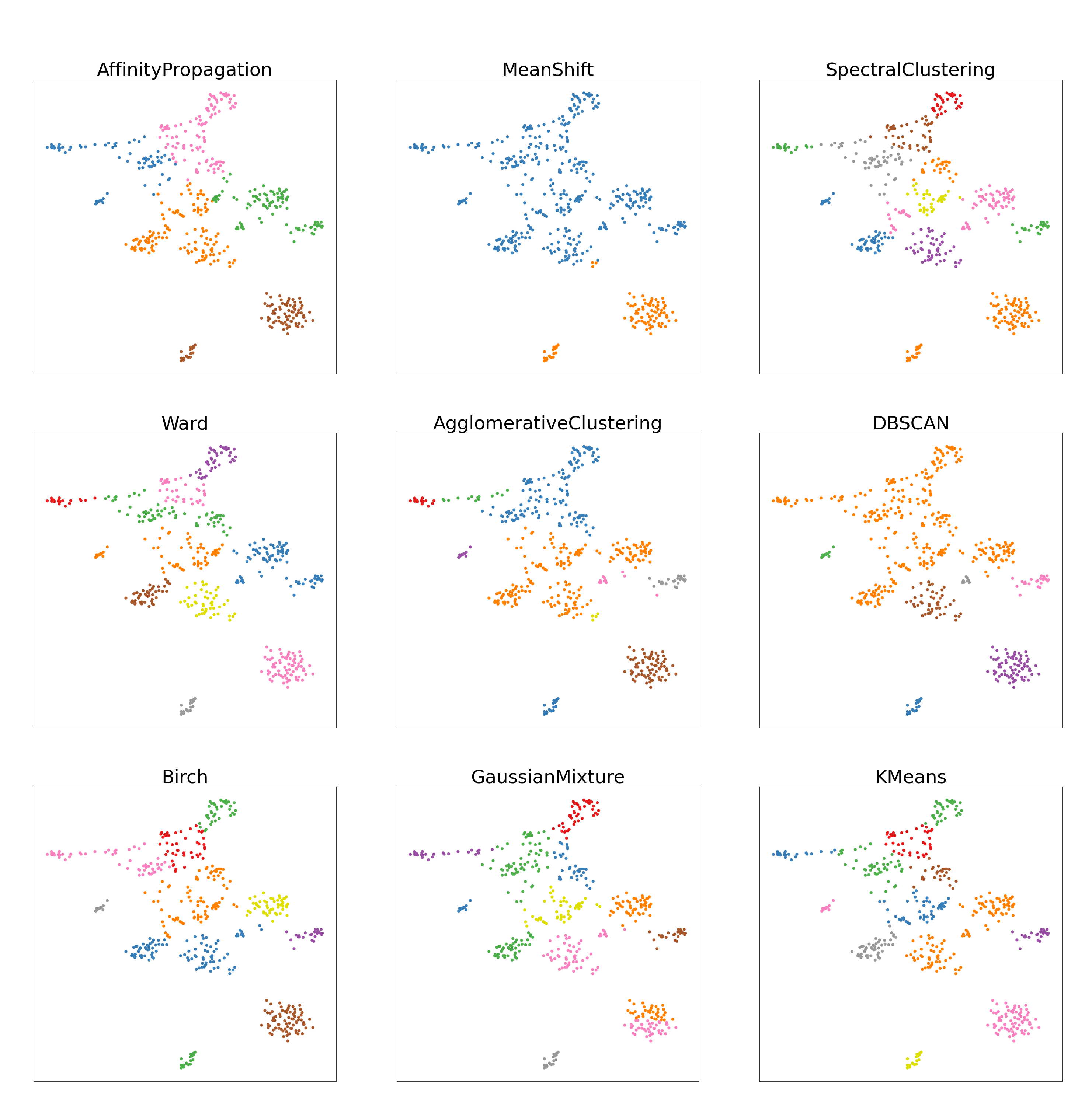}
    \caption{Patient ID}
    \label{acpatient}
\end{subfigure}%
\caption{Model Comparisons}
\end{figure}
%
\indent To review, the goal for molecular subtype was to separate the entire dataset into four subpopulations (estrogen receptor positive, estrogen receptor positive, double positive, triple negative breast cancer, and human epidermal growth factor). A satisfactory job visually is a graph where four distinct clusters are seen. Figure \ref{fig:acmol} shows that Affinity Propagation, Agglomerative Clustering, Mean Shift, and DBSCAN have large differences from the rest of the models. Affinity Propagation and Mean Shift have a different number of clusters while Agglomerative and DBSCAN have differently sized clusters than most clustering algorithms. The rest of the models (Spectral, Ward, BIRCH, Gaussian Mixture Model, and KMeans) have similar structure with the same number of clusters and roughly the same partitions. \\
\indent In the same dataset, the data can also be separated according to cell state (tumor versus nonTumor) which results in two subpopulations. For tumor cells, Figure \ref{fig:actumor} Affinity Propagation and DBSCAN are extremely different from the rest of the models, for they identify a significantly larger number of clusters. BIRCH and KMeans have roughy the same partitions while the rest of the models (Spectral, Ward, Agglomerative, Gaussian Mixture Model, and Mean Shift) also have the same partitions. Every model except Affinity Propagation and DBSCAN created two clusters. Because there were only two groups to cluster, the consistency of partitioning for the clustering algorithms was higher than with molecular subtype and patient ID. \\
\indent The third goal is to separate the entire dataset into thirteen distinct clusters representing different patient identifications. For patient ID, Figure \ref{acpatient} reveals that DBSCAN, Agglomerative, and Mean Shift clustering all have distinct different partitioning from the others models in that they create one large cluster and the rest are significantly smaller. The number of clusters is also somewhat different from the rest of the models. The other models all have roughly the same partitioning and size in clusters. With thirteen different groups, there is a lot more variance with partitioning and creating groups. 
\subsection{Evaluation Algorithm}
\indent In the following calculations, the columns of each matrix represent the percentage values of each cluster. The rows of each matrix represent the percentage values of each subtype. Therefore, in the following matrix A, $A_{i, j}$ is the percentage of subtype i in cluster j. For example, in equation 24, $A_{1, 1}$ holds the value 1.0 representing 100\% of $C_2$ (cluster 2) is comprised of subtype ERPCELL. In the equations, the crossed out entries represent the disregarded percentages resulting from their elimination during the evaluation algorithm; an entry is crossed out if the next absolute maximum value remaining in the matrix is in the same row or column.  
\begin{figure}[H]
\begin{subfigure}{0.5\textwidth}
\caption{Accuracy Calculation Ideal Case}
\begin{equation}
\begin{blockarray}{ccccc}
 & C_1 & C_2 & C_3 & C_4\\
\begin{block}{c(cccc)}
\text{DPCELL} & 1.0 & 0.0 & 0.0 & 0.0\\
\text{ERPCELL} & 0.0 & 1.0 & 0.0 & 0.0\\
\text{HER2P} & 0.0 & 0.0 & 1.0 & 0.0\\
\text{TNBC} & 0.0 & 0.0 & 0.0 & 1.0\\
\end{block}
\end{blockarray}
\end{equation}
\vspace{0.2cm}\\
\begin{equation}
\begin{blockarray}{ccccc}
& C_1 & C_2 & C_3 & C_4\\
\begin{block}{c(cccc)}
\text{DPCELL} & 1.0 & \xcancel{0.0} & \xcancel{0.0} & \xcancel{0.0}\\
\text{ERPCELL} & \xcancel{0.0} & 1.0 & \xcancel{0.0} & \xcancel {0.0}\\
\text{HER2P} & \xcancel{0.0} & \xcancel{0.0} & 1.0 & \xcancel{0.0}\\
\text{TNBC} & \xcancel{0.0} & \xcancel{0.0} & \xcancel{0.0} & 1.0\\
\end{block}
\end{blockarray}
\end{equation}
\vspace{0.2cm}\\
\begin{equation}
\begin{aligned}
C_1 &= \text{DPCELL } (1.0) \\
C_2 &= \text{ERPCELL } (1.0) \\
C_3 &= \text{HER2P } (1.0) \\
C_4 &= \text{TNBC } (1.0)\\
\end{aligned}
\end{equation}
\end{subfigure}
\begin{subfigure}{0.5\textwidth}
\caption{Accuracy Calculation Non-Ideal Case}
\begin{equation}
\begin{blockarray}{ccccc}
 & C_1 & C_2 & C_3 & C_4\\
\begin{block}{c(cccc)}
\text{DPCELL} & 0.74 & 0.2 & 0.1 & 0.1\\
\text{ERPCELL} & 0.2 & 0.15 & 0.3 & 0.5\\
\text{HER2P} & 0.03 & 0.2 & 0.6 & 0.4\\
\text{TNBC} & 0.03 & 0.55 & 0.0 & 0.1\\
\end{block}
\end{blockarray}
\end{equation}
\vspace{0.2cm}\\
\begin{equation}
\begin{blockarray}{ccccc}
& C_1 & C_2 & C_3 & C_4\\
\begin{block}{c(cccc)}
\text{DPCELL} & 0.74 & \xcancel{0.2} & \xcancel{0.1} & \xcancel{0.1}\\
\text{ERPCELL} & \xcancel{0.2} & \xcancel{0.15} & \xcancel{0.3} & 0.5\\
\text{HER2P} & \xcancel{0.03} & \xcancel{0.2} & 0.6 & \xcancel{0.4}\\
\text{TNBC} & \xcancel{0.03} & 0.55 & \xcancel{0.0} & \xcancel{0.1}\\
\end{block}
\end{blockarray}
\end{equation}
\vspace{0.2cm}\\
\begin{equation}
\begin{aligned}
C_1 &= \text{DPCELL } (0.74) \\
C_2 &= \text{TNBC } (0.55) \\
C_3 &= \text{HER2P } (0.6) \\
C_4 &= \text{ERPCELL } (0.5)\\
\end{aligned}
\end{equation}

\end{subfigure}
 \end{figure}
 

\indent To evaluate the accuracy of the models in this work, the truth dataset is examined and files are created containing each "true" population or what each cluster should be made of. Then, the cells are compared to every true cell to see what their true label should be. The percentages of what cells (ERPCELL, DPCELL, HER2P, and TNBC), (tumor v. nonTumor), or (BC01, BC02, BC03, BC04, BC05, BC06, BC07, BC08, BC09, BC10, BC11, BC03\_LN, BC07\_LN) are represented in each cluster will be written to files. Then, the assumption that each cluster is supposed to represent a singular population will be made and each cell population that is the majority of each cluster is marked as correct whereas the remaining cells of that cluster are marked incorrect. The accuracy is represented as the number of "correct" cells over the total number of cells clustered. \\
\indent To validate this method of evaluation, the truth data was completely randomized and partially randomized (10\% of the true labels were mixed up); then 100 simulations were ran. If the projected accuracies showed a vast difference with the completely randomized validation set and a slight difference with the partially randomized validation set, then the method for evaluating the algorithms can be deemed efficient. 
         

\section{Results}
We start by exhibiting the visualizations produced through the evaluation algorithm compared to the biological truth of each clustering algorithm. The visualizations are then followed by a brief discussion of the observations. The analysis of 100 simulations is then discussed and followed by an explanation of methods used for validation of the evaluation algorithm. 
\subsection{Visualizations}
\begin{figure}[H]
\begin{subfigure}{0.5\textwidth}
    \centering
    \includegraphics[width=0.9\textwidth]{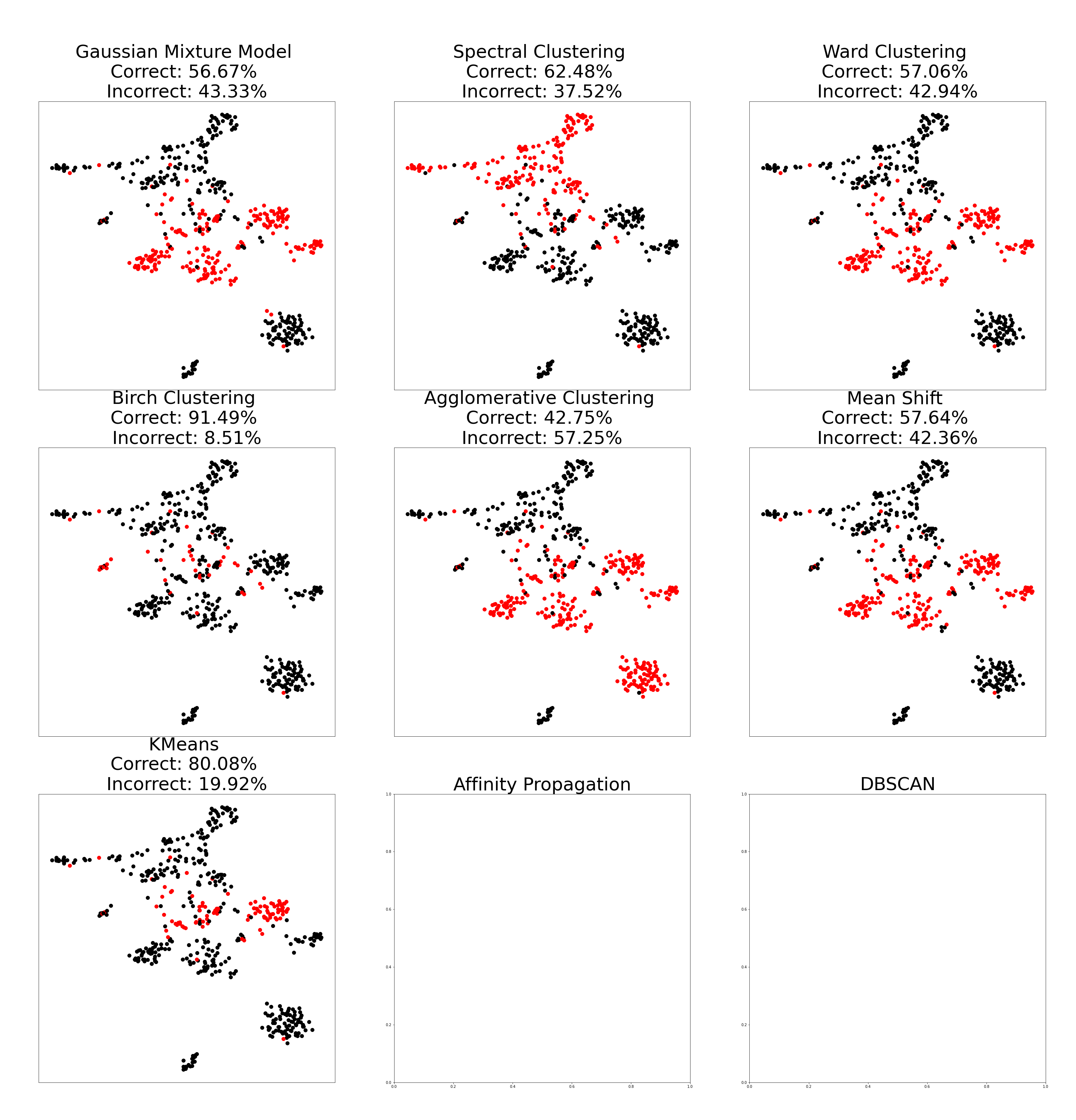}
    \caption{Truth v. Predicted}
\end{subfigure}%
\begin{subfigure}{0.5\textwidth}
    \centering
    \includegraphics[width=0.9\textwidth]{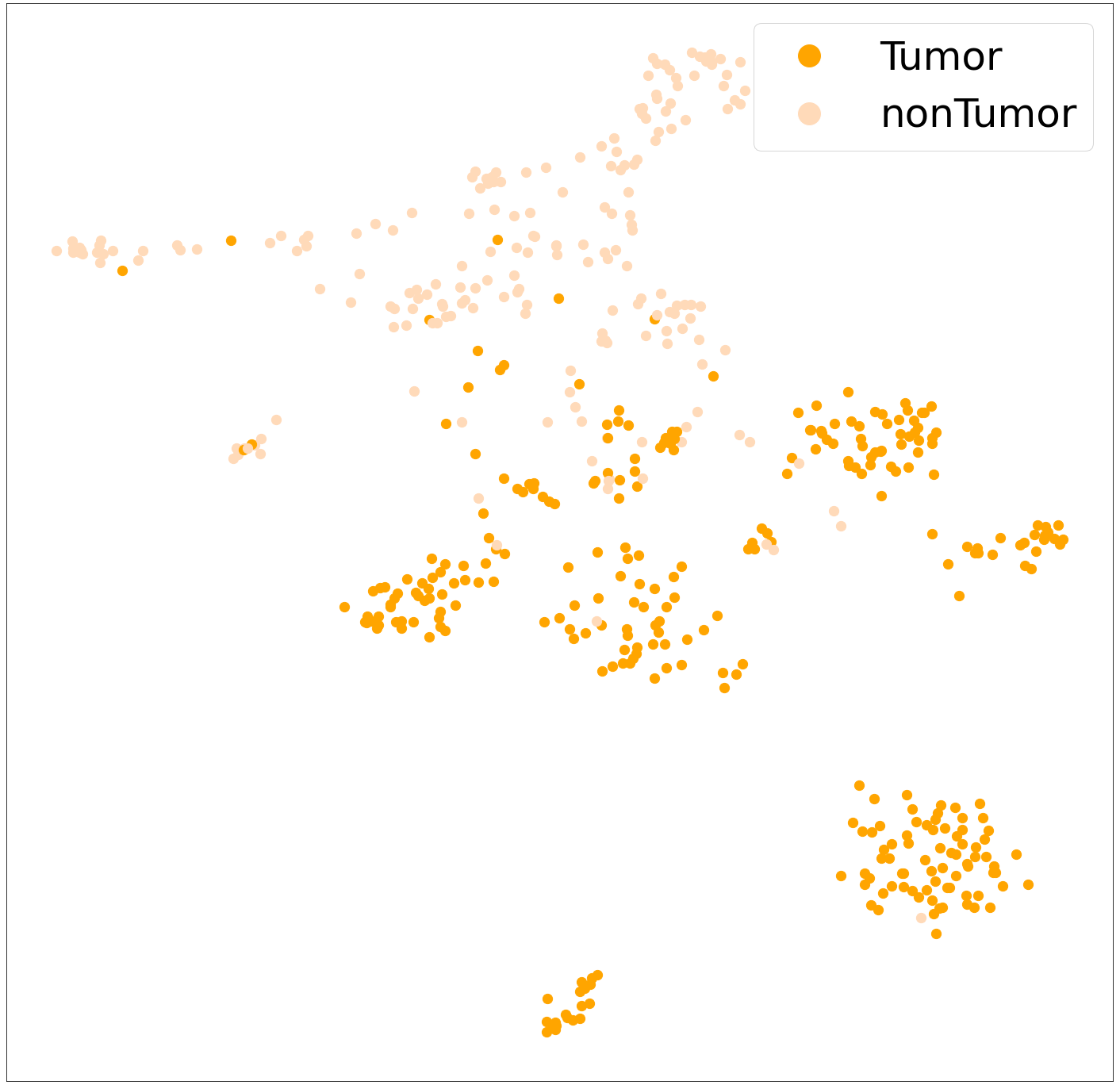}
     \caption{True Tumor Cell Diagram}
\end{subfigure}
\caption{Tumor Plots}
\label{tumorviz}
\end{figure}
\begin{figure}[H]
\begin{subfigure}{0.5\textwidth}
    \centering
    \includegraphics[width=0.9\textwidth]{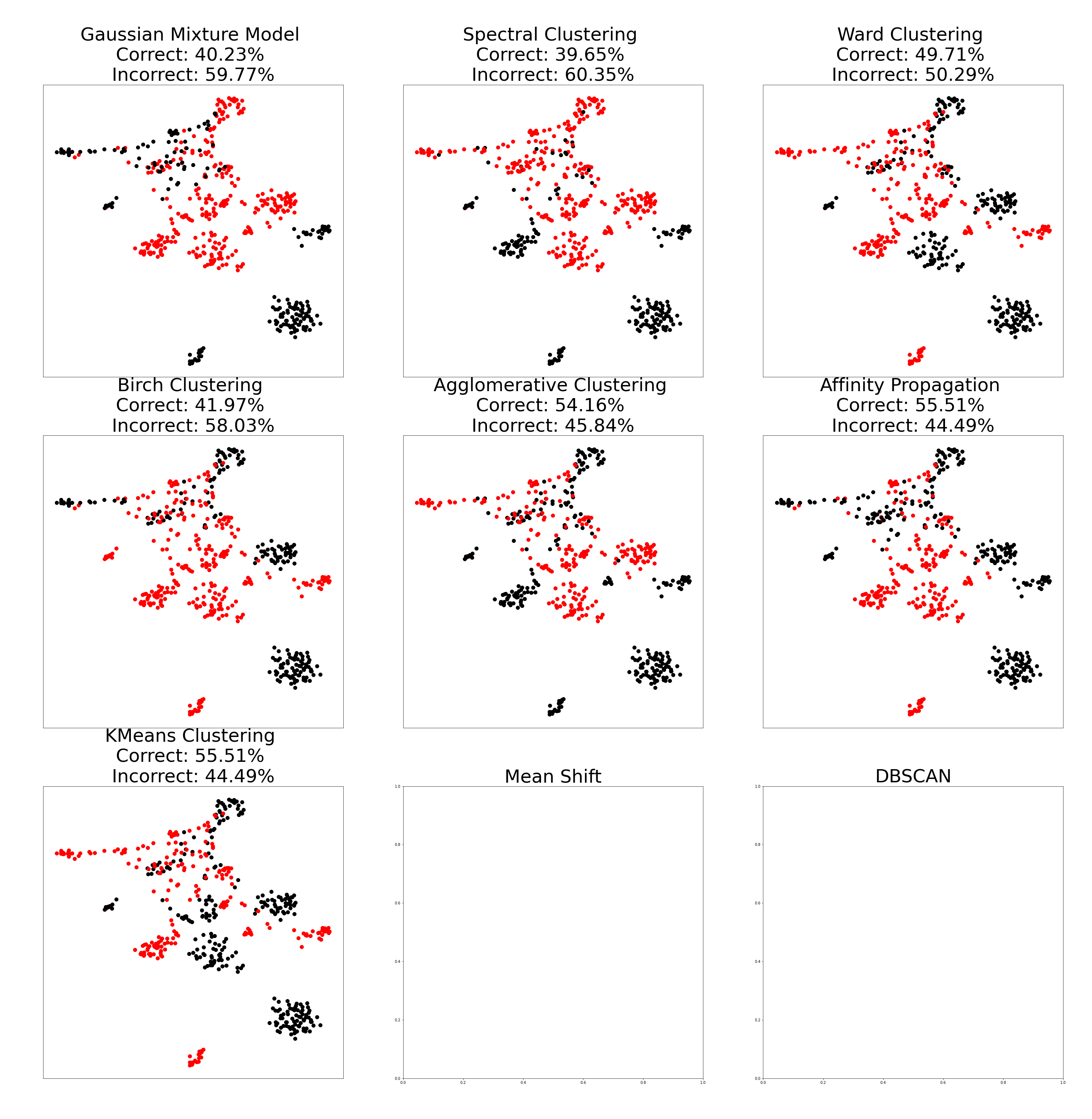}
    \caption{Truth v. Predicted}
\end{subfigure}%
\begin{subfigure}{0.5\textwidth}
    \centering
    \includegraphics[width=0.9\textwidth]{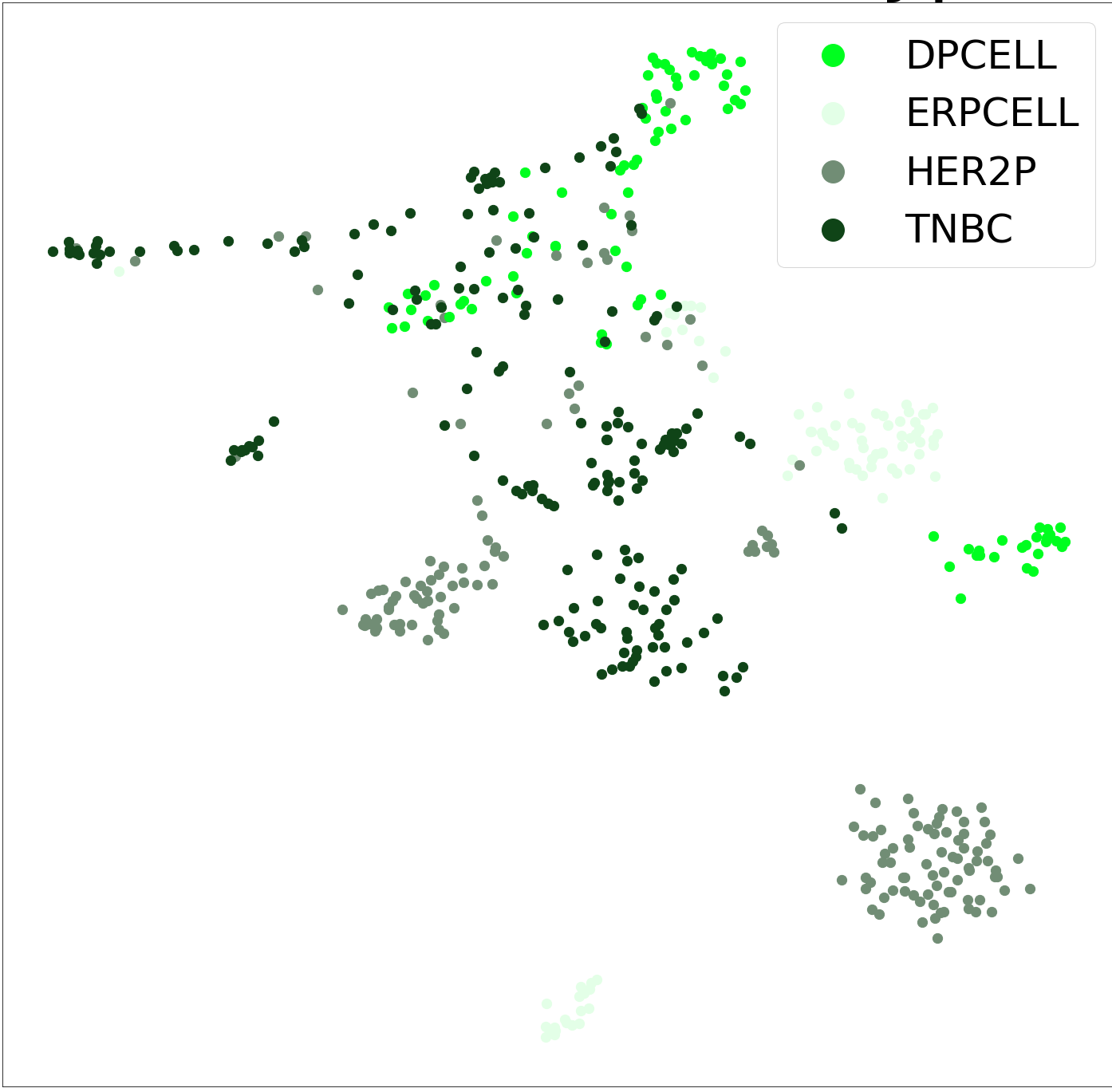}
     \caption{True Molecular Subtype Diagram}
\end{subfigure}
\caption{Molecular Subtype Plots}
\label{molviz}
\end{figure}

\begin{figure}[H]
\begin{subfigure}{0.5\textwidth}
    \centering
    \includegraphics[width=0.9\textwidth]{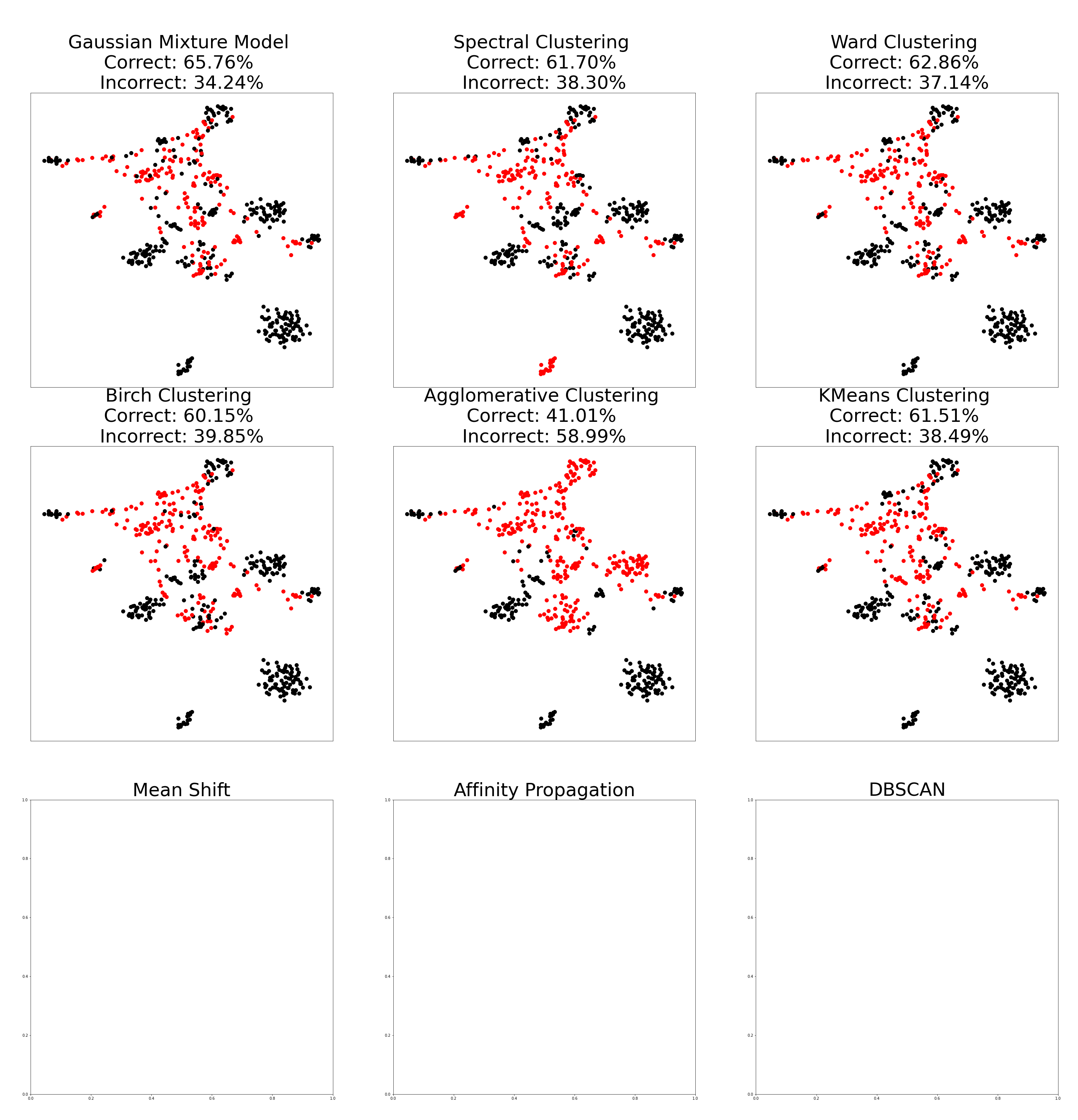}
    \caption{Truth v. Predicted}
\end{subfigure}%
\begin{subfigure}{0.5\textwidth}
    \centering
    \includegraphics[width=0.9\textwidth]{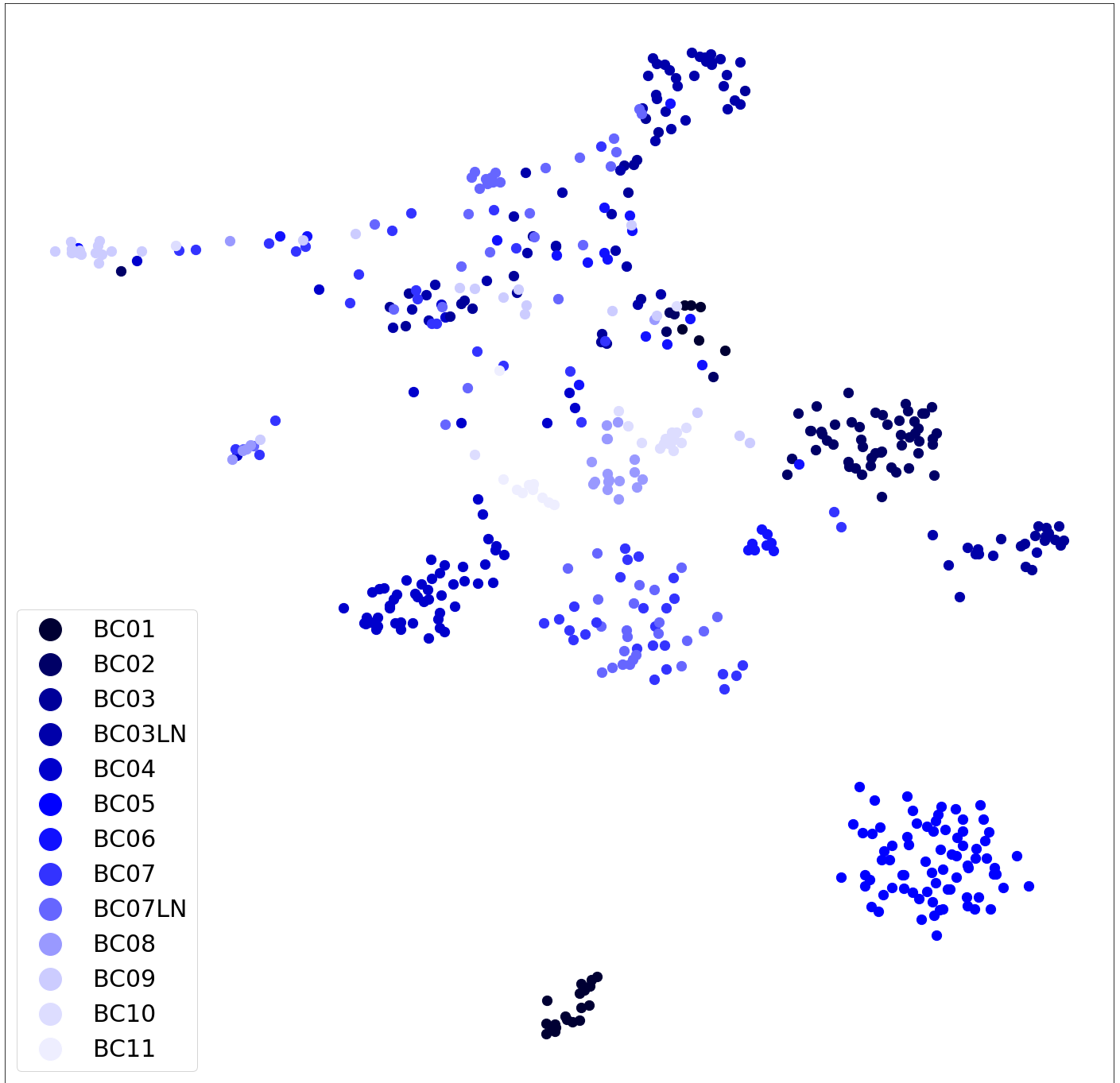}
    \caption{True Patient ID Diagram}
\end{subfigure}
\caption{Patient ID Plots}
\label{patientviz}
\end{figure}
\indent For each of the figures, the red points represent the cells that were incorrectly clustered, and the black points represent the cells that were correctly clustered. The evaluation algorithm gauged the homogeneity of each cluster where the cells within the cluster are of the same type. With comparison to the truth data, the evaluation algorithm gauges the percentage of each subtype represented in each cluster and iteratively mark which cluster represented which subpopulation based on where the highest concentration of each subtype was. The red data points represent outliers, inconsistent points that were grouped with dissimilar cells. The black data points represent homogeneous subgroups that were correctly clustered. 
\indent For the tumor plots (Fig. \ref{tumorviz}), it is visually apparent that for this single run BIRCH and KMeans clustering performed the best with accuracy scores above 80\%, with errors primarily on the line of partition between the clusters. The rest of the models were significantly off with their areas of partition which resulted in a lower accuracy score. Most models were able to group together the tumor cells correctly excluding Spectral clustering. \\
\indent For the molecular subtype plots (Fig. \ref{molviz}), the models performed significantly worse than they did on clustering the tumor cells. The highest performing models for this specific run were KMeans, Agglomerative, and Affinity Propagation clustering. Agglomerative and Affinity Propagation clustering are both very low performing for the other tasks. Regardless, these scores of approximately 55\% are vastly lower than the previous scores of above 80\% for partitioning the tumor cells. From the visualizations, it is apparent that none of the models did a satisfactory job of accurately clustering the cells. \\
\indent For the patient ID plots (Fig. \ref{patientviz}), all of the models scored between 60\% to 65\% accuracy excluding Agglomerative clustering. A threshold for accuracy of an unsupervised algorithm should be around 80\% accuracy because that accounts for most of the data. Considering even the highest performing models scored well below that set threshold, none of them did a strong job of clustering these cells for patient ID. 
\subsection{Analysis}
\begin{figure}[H]
    \begin{subfigure}{0.5\textwidth}
    \centering
    \includegraphics[width=0.9\textwidth]{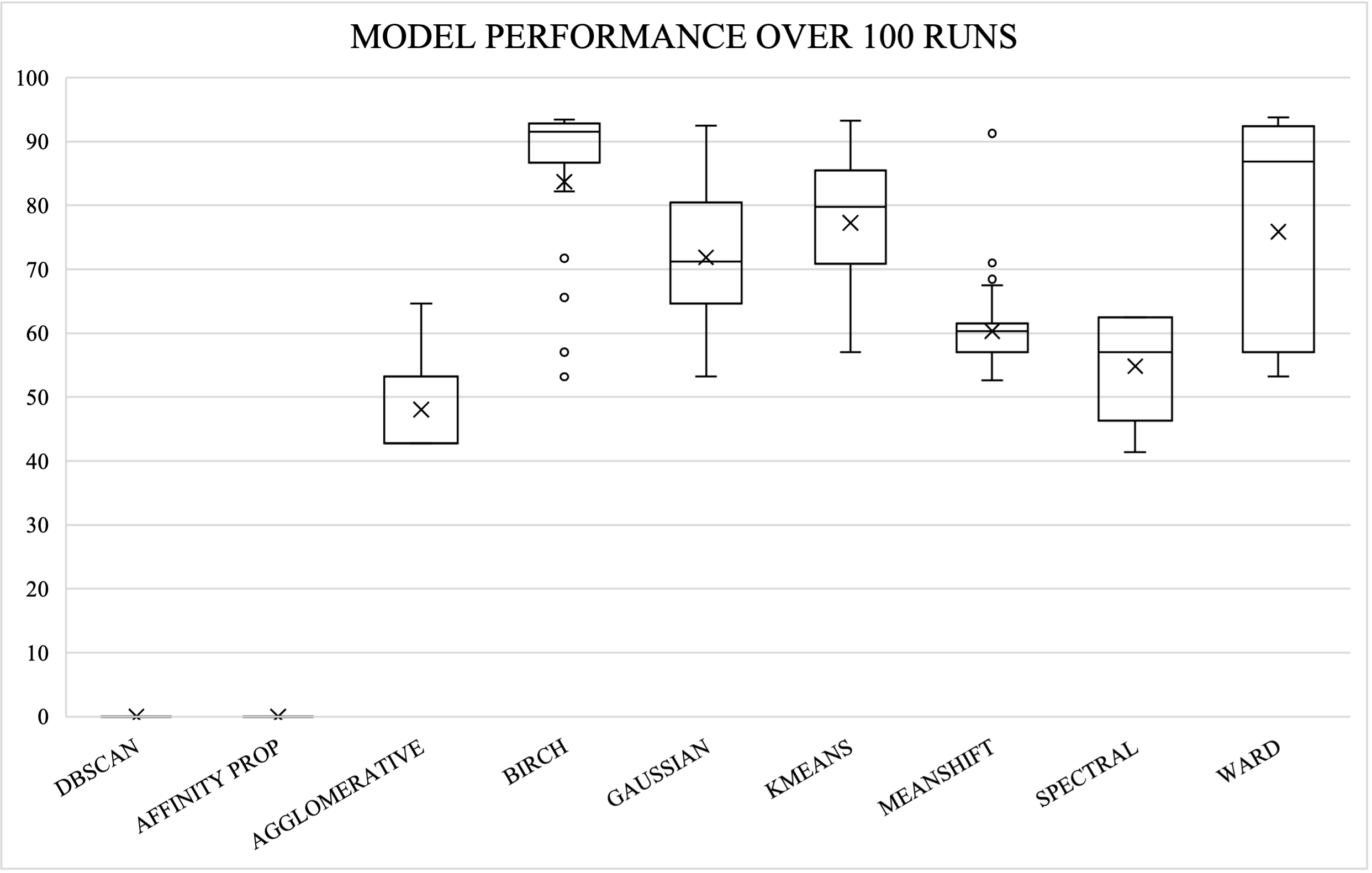}
    \end{subfigure}
    \begin{subfigure}{0.5\textwidth}
    \centering
    \includegraphics[width=0.9\textwidth]{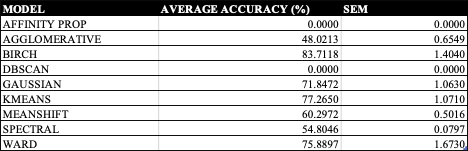}
    \end{subfigure}
    \caption{Tumor Cell Plots}
    \label{tumoreval}
\end{figure}
\begin{figure}[H]
    \begin{subfigure}{0.5\textwidth}
    \centering
    \includegraphics[width=0.9\textwidth]{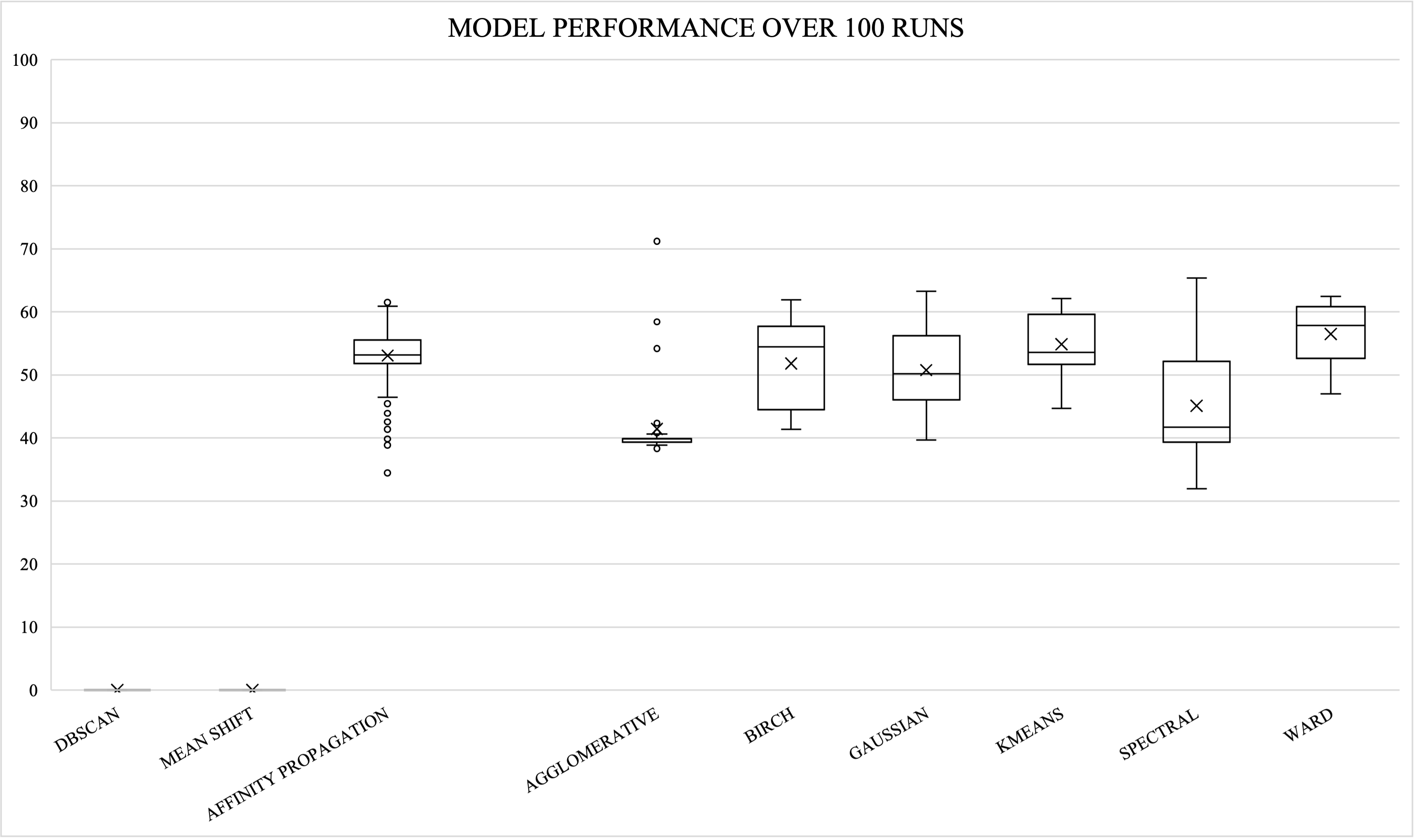}
    \end{subfigure}
    \begin{subfigure}{0.5\textwidth}
    \centering
    \includegraphics[width=0.9\textwidth]{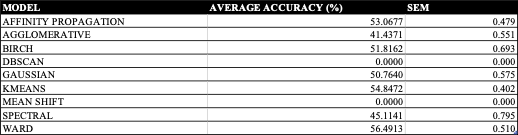}
    \end{subfigure}
    \caption{Molecular Subtype Plots}
    \label{moleval}
\end{figure}
\begin{figure}[H]
    \begin{subfigure}{0.5\textwidth}
    \centering
    \includegraphics[width=0.9\textwidth]{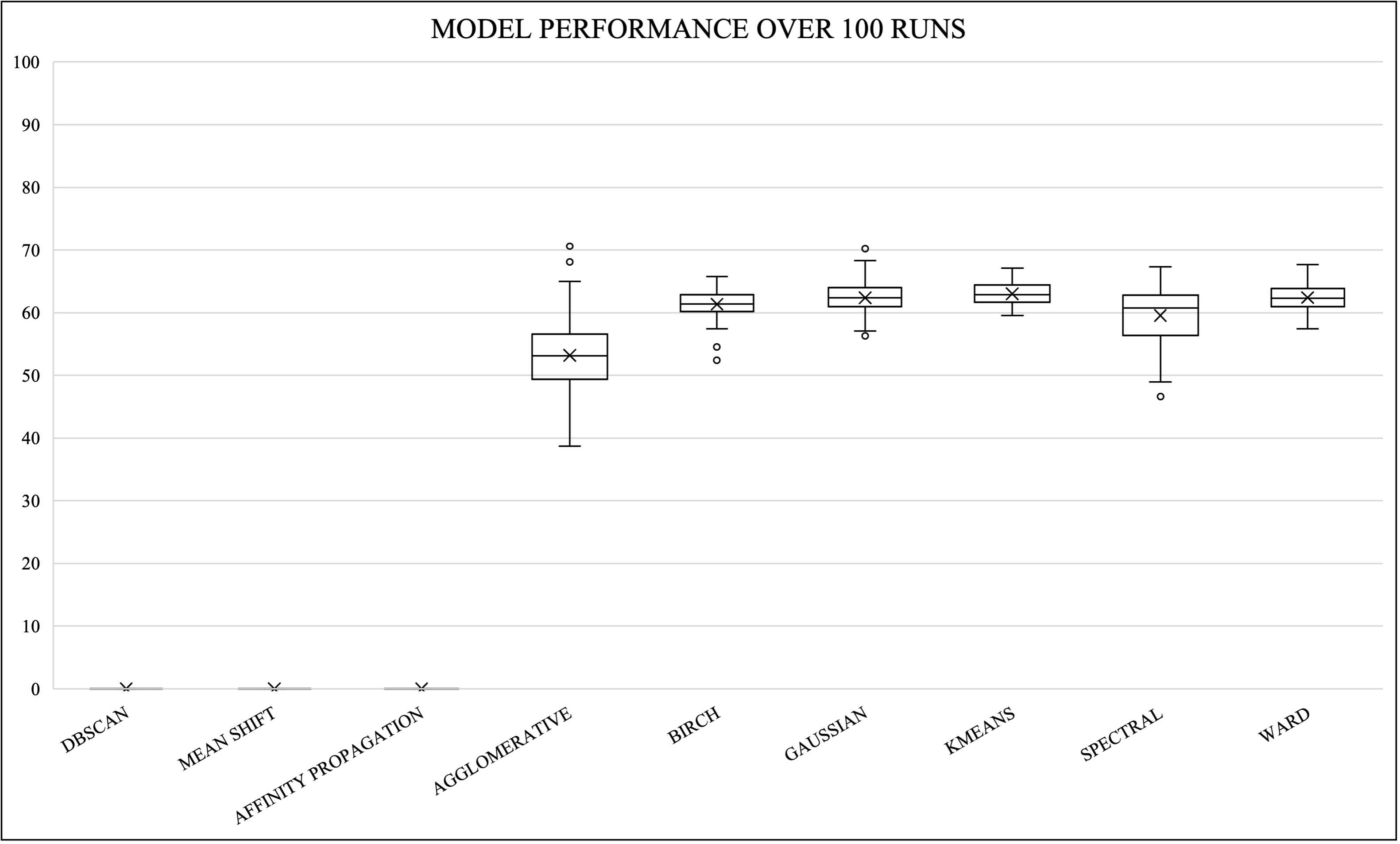} 
    \end{subfigure}
    \begin{subfigure}{0.5\textwidth}
    \centering
    \includegraphics[width=0.9\textwidth]{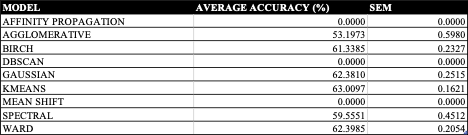}
    \end{subfigure}
    \caption{Patient ID Plots}
    \label{patienteval}
\end{figure}
 \indent For the tumor cell evaluation (Fig. \ref{tumoreval}), the highest performing models were BIRCH (83.7\%), KMeans (77.3\%), and Ward(75.9\%) clustering. All of these models performed relatively close to the threshold of 80\%. The lowest performing model that generated the correct number of clusters was Agglomerative clustering (48.0\%). DBSCAN and Affinity Propagation failed to separate the total population into the correct number of subpopulations. The models with the most variation were Ward (1.67\% Standard Error of the Mean (SEM)) and BIRCH (1.40\% SEM), and the models with the least variation were Spectral (0.080\% SEM) and MeanShift (0.502\% SEM).  On criteria of high mean accuracy and relatively low variation, KMeans performed well having the second highest accuracy and a median SEM. BIRCH had a vastly higher accuracy than any other model which also qualifies it as a high performing model for the task of classifying tumor cell populations. \\
\indent For the molecular subtype evaluation (Fig. \ref{moleval}), the highest performing models were Ward (56.5\%), KMeans (54.8\%), and Affinity Propagation (53.1\%). The lowest performing models that generated the correct number of clusters were Spectral (45.1\%) and Agglomerative (41.4\%) clustering. DBSCAN and MeanShift failed to generate the correct number of subpopulations. The models with the highest variation over 100 runs were Spectral (0.795\% SEM) and BIRCH (0.693\% SEM); the models with the lowest variation over 100 runs were Affinity Propagation (0.479\% SEM) and Kmeans (0.402\%). On the criteria of high mean accuracy and relatively low variation, both KMeans and Affinity Propagation performed best with the highest accuracies and lowest SEM. \\
\indent For the patient ID evaluation (Fig. \ref{patienteval}), the highest performing models were KMeans (63.0\%), Ward (62.4\%), Gaussian Mixture Model (62.4\%), and BIRCH (61.34\%). Affinity Propagation, DBSCAN, and Mean Shift failed to generate the correct number of subpopulations. The lowest performing models were Agglomerative (53.2\%) and Spectral (59.6\%). The models with the most variation over 100 runs were Agglomerative (0.598\% SEM) and Spectral (0.451\%); the models with the lowest variation were KMeans (0.162\% SEM), Ward (0.205\% SEM), and BIRCH (0.233\% SEM). On the criteria of high mean accuracy and low variation, KMeans, Ward, and BIRCH clustering are considered the best performing models. 
 \subsection{Validation}
 \begin{figure}[H]
    \begin{subfigure}{0.5\textwidth}
     \centering
     \includegraphics[width=0.9\textwidth]{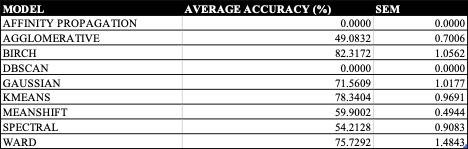}
     \caption{Partial Randomization Evaluation}
     \end{subfigure}
     \begin{subfigure}{0.5\textwidth}
     \centering
     \includegraphics[width=0.9\textwidth]{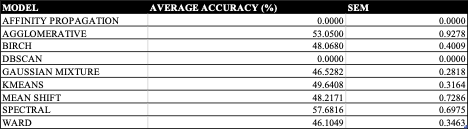}
     \caption{Complete Randomization Evaluation}
     \end{subfigure}
     \caption{Tumor Cell Clustering Evaluation}
 \end{figure}
 \begin{figure}[H]
    \begin{subfigure}{0.5\textwidth}
     \centering
     \includegraphics[width=0.9\textwidth]{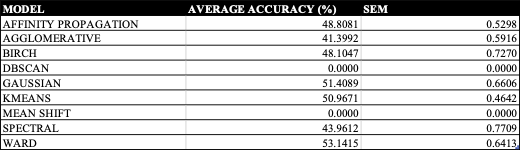}
     \caption{Partial Randomization Evaluation}
     \end{subfigure}
     \begin{subfigure}{0.5\textwidth}
     \centering
     \includegraphics[width=0.9\textwidth]{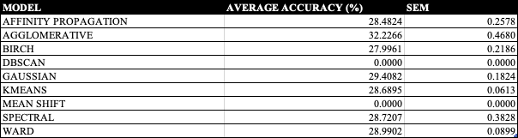}
     \caption{Complete Randomization Evaluation}
     \end{subfigure}
     \caption{Molecular Subtype Cell Clustering Evaluation}
 \end{figure}
 \begin{figure}[H]
    \begin{subfigure}{0.5\textwidth}
     \centering
     \includegraphics[width=0.9\textwidth]{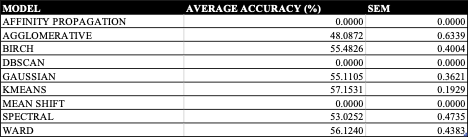}
     \caption{Partial Randomization Evaluation}
     \end{subfigure}
     \begin{subfigure}{0.5\textwidth}
     \centering
     \includegraphics[width=0.9\textwidth]{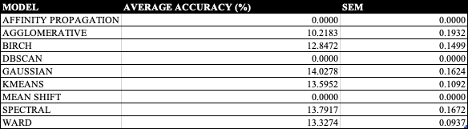}
     \caption{Complete Randomization Evaluation}
     \end{subfigure}
     \caption{Patient ID Cell Clustering Evaluation}
 \end{figure}
 These randomization tests serve as a method to validate the method of evaluation for the algorithms. Ideally, a robust algorithm would be sensitive to changes in the truth datasets. With partial randomization, the accuracy scores for each model should drop slightly compared to no randomization at all. With complete randomization, the accuracy scores should drop largely compared to no randomization. \\
 \indent In this case, randomization refers to randomizing the truth labels themselves by assigning each cell to a different label than the true label. The evaluation algorithm is expected to assign lower scores than initially calculated because the truth data is completely different than the model predictions. However, it is likely each accuracy score will still be above 0\% for complete randomization because there is a chance that after randomization some cells still have their true labels; there would just be a significantly fewer number of them.\\
 \indent For each task, complete randomization did result in significantly lower scores. Generally, if a model scored approximately 60\% under no randomization, complete randomization would generate a score of approximately 10\%. Again for each task, partial randomization showed a slightly lower range of scores for each model with a drop in scores of about 5\%-10\% for each model. 
\section{Discussion}
\indent We investigated the efficacy of unsupervised analysis on scRNAseq data through a pipeline and evaluation metrics to rank nine models (BIRCH, Ward, Spectral, Gaussian Mixture Model, DBSCAN, Affinity Propagation, Agglomerative Average Linkage, Mean Shift, and K-Means). With one dataset, we sought to evaluate these models on their ability to separate subclonal cell populations represented within the data. These subpopulations were cell state (tumor versus nonTumor), molecular subtype (estrogen receptor positive, estrogen receptor positive, double positive, triple negative breast cancer, and human epidermal growth factor), and patient identification(BC01, BC02, BC03, BC03LN, BC04, BC05, BC06, BC07, BC07LN, BC08, BC09, BC10, BC11). \\
\indent In this research, unsupervised learning is deemed effective in single cell RNA sequencing data projection and analysis since clusters were created that were both consistent and successful. Most models would effectively distinguish the correct number of subpopulations with the exception of DBSCAN, Affinity Propagation, and Mean Shift for multiple tasks. Of these models, KMeans, BIRCH, and Ward clustering have the highest mean accuracy and low variation over 100 simulations for most tasks. BIRCH excelled at classification of tumor cell data with an average accuracy above 80\% while KMeans and Ward performed best on both patient ID and molecular subtype classification. Dimensionality reduction with UMAP and PCA is an effective technique for reducing the number of components in the dataset to effectively cluster the data points. By previewing the biological mappings, it can be seen that biological structure preserved in that distinct partitions between subtypes were created. \\
\indent BIRCH scoring above 80\% on average for accuracy in clustering tumor cell data shows that it is effective in clustering tumor v. nontumor cells. However, for the rest of the tasks (patient ID and molecular subtype), the models often performed at an accuracy of about 60\% which is below the typical performance target. Of these models, KMeans and Ward were the highest performing. Their high performance, however, does not mean that these should be used as is for the tasks specifically. Often, the method that Ward and KMeans have for partitioning the data would respond best to the structure, yet there is still a large room for improvement. This can also be an indication that gene expression levels are not the best way of classfying molecular subtype or patient ID, for the models were not able to recognize distinct differences in the levels to strongly cluster the data. An optimized pipeline using BIRCH, KMeans, and Ward clustering for each task along with proper parameterization for UMAP and PCA would prove to be as effective as possible for classifying cell subpopulations. Although not entirely accurate, these were the highest performing models from the simulations that show potential to generate proper clusters of tumor v. nonTumor cells, patient ID, and molecular subtype which may help recognize different phenotypes within cell populations. \\
\indent An evaluation of the most frequently used models is particularly important, for there has not been an in depth discussion of how these unsupervised methods compare to each other for scRNAseq data analysis. In a clinical setting, this optimized pipeline can be used to improve scRNAseq analysis because it incorporates the most efficient and effective techniques found in this study. By using the pipeline, clinicians can learn more about the nature of malignant cell growth and tumor heterogeneity which directly affects the quality of cancer treatment.  
\section{Acknowledgements}
This work is supported by the NIH NCI specifically from the grant NIH-CSBC: U54 CA209978. I would like to thank my mentor Dr. Frederick R. Adler at the University of Utah Math Department, and Dr. Jim Waisman and team for the supply of clinical data. I would also like to thank Dr. Jeffrey Chang and his group for assembling the data and Dr. Mark Smithson for help during the initial phases of this work.

\bibliographystyle{unsrt}
\bibliography{main}

\begin{thebibliography}{10}

\bibitem{proposal}
Jasmine~A. McQuerry, Jeffrey~T. Chang, David D.~L. Bowtell, Adam Cohen, and
  Andrea~H. Bild.
\newblock Mechanisms and clinical implications of tumor heterogeneity and
  convergence on recurrent phenotypes.
\newblock {\em Journal of Molecular Medicine}, 95(11):1167--1178, Nov 2017.

\bibitem{intra_tumor}
Giorgio Stanta and Serena Bonin.
\newblock Overview on clinical relevance of intra-tumor heterogeneity.
\newblock {\em Frontiers in Medicine}, 5:85, 2018.

\bibitem{targetted_therapy}
L.~Yan, N.~Rosen, and C.~Arteaga.
\newblock {{T}argeted cancer therapies}.
\newblock {\em Chin J Cancer}, 30(1):1--4, Jan 2011.

\bibitem{sc_rna_tech}
B.~Hwang, J.~H. Lee, and D.~Bang.
\newblock {{S}ingle-cell {R}{N}{A} sequencing technologies and bioinformatics
  pipelines}.
\newblock {\em Exp Mol Med}, 50(8):1--14, 08 2018.

\bibitem{single_cell_meets_once_again}
S.~Kolchenko.
\newblock Machine learning meets biology (once again) or single cell rna
  sequencing as a field for unsupervised learning.
\newblock
  https://sergeykolchenko.medium.com/machine-learning-meets-biology-once-again-or-single-cell-rna-sequencing-as-a-field-for-2be24f9108dd,
  2019.

\bibitem{birch_cluster}
Alakesh Bora.
\newblock Ml birch clustering.
\newblock https://www.geeksforgeeks.org/ml-birch-clustering/, July 2020.

\bibitem{BIRCH_paper}
Tian Zhang, Raghu Ramakrishnan, and Miron Livny.
\newblock Birch: An efficient data clustering method for very large databases.
\newblock {\em SIGMOD Rec.}, 25(2):103–114, June 1996.

\bibitem{ward_penn}
14.7 - ward's method: Stat 505.
\newblock https://online.stat.psu.edu/stat505/lesson/14/14.7.

\bibitem{ward_method}
Ward's method (minimum variance method).
\newblock https://www.statisticshowto.com/wards-method/, December 2020.

\bibitem{spectral_foundation}
William Fleshman.
\newblock Spectral clustering: Foundation and application.
\newblock https://towardsdatascience.com/spectral-clustering-aba2640c0d5b,
  February 2019.

\bibitem{spectral_overview}
Neerja Doshi.
\newblock Spectral clustering: The math and intuition behind how it works.
\newblock https://towardsdatascience.com/spectral-clustering-82d3cff3d3b7,
  February 2019.

\bibitem{gmm}
Oscar~Contreras Carrasco.
\newblock Gaussian mixture models explained.
\newblock
  https://towardsdatascience.com/gaussian-mixture-models-explained-6986aaf5a95,
  June 2019.

\bibitem{gmm2}
John McGonagle, Geoff Pilling, and Andre Dobre.
\newblock Gaussian mixture model.
\newblock https://brilliant.org/wiki/gaussian-mixture-model/.

\bibitem{better_gmm}
Aishwarya Singh.
\newblock Build better and accurate clusters with gaussian mixture models.
\newblock
  https://www.analyticsvidhya.com/blog/2019/10/gaussian-mixture-models-clustering/,
  2019.

\bibitem{dbscan_theory}
Jared Gerschler.
\newblock The dbscan algorithm background and theory.
\newblock
  https://jaredgerschler.blog/2017/01/13/the-dbscan-algorithm-background-and-theory/,
  2017.

\bibitem{dbscan_math}
Dbscan with math.
\newblock
  https://medium.com/@rdhawan201455/dbscan-clustering-algorithm-with-maths-c40dcee88281,
  November 2019.

\bibitem{aff_prop_math}
Harshita Vemula.
\newblock How affinity propagation works.
\newblock
  https://towardsdatascience.com/math-and-intuition-behind-affinity-propagation-4ec5feae5b23,
  August 2019.

\bibitem{agglomerative_hierarchical}
Agglomerative hierarchical clustering.
\newblock
  https://www.datanovia.com/en/lessons/agglomerative-hierarchical-clustering/.

\bibitem{agglom}
Dr.~Saed Sayad.
\newblock Hierarchical clustering.
\newblock https://www.saedsayad.com/clustering\_hierarchical.htm, 2021.

\bibitem{stanford_avg_link}
Cambridge University.
\newblock Single-link, complete-link \& average-link clustering.
\newblock https://nlp.stanford.edu/IR\-book/completelink.html, 2008.

\bibitem{spin_ms}
Matt Nedrich.
\newblock Mean shift clustering.
\newblock https://spin.atomicobject.com/2015/05/26/mean-shift-clustering/,
  2015.

\bibitem{ms_eq}
Miguel~{\'A}. Carreira-Perpi{\~{n}}{\'a}n and Christopher K.~I. Williams.
\newblock On the number of modes of a gaussian mixture.
\newblock pages 625--640, 2003.

\bibitem{gg}
Mean-shift clustering.
\newblock https://www.geeksforgeeks.org/ml-mean-shift-clustering/, May 2019.

\bibitem{k_medium}
Madison Schott.
\newblock K-means clustering algorithm for machine learning.
\newblock
  https://medium.com/capital-one-tech/k-means-clustering-algorithm-for-machine-learning-d1d7dc5de882,
  April 2019.

\bibitem{k_understanding}
Nikita Sharma.
\newblock Understanding the mathematics behind k-means clustering.
\newblock
  https://heartbeat.fritz.ai/understanding-the-mathematics-behind-k-means-clustering-40e1d55e2f4c?gi=d3462e3dc02f,
  February 2020.

\bibitem{onestop_pca}
M.~Brems.
\newblock A one-stop shop for principal component analysis.
\newblock
  https://towardsdatascience.com/a-one-stop-shop-for-principal-component-analysis-5582fb7e0a9c,
  2019.

\bibitem{interpret_pca}
Interpret the key results for principal components analysis.
\newblock
  https://support.minitab.com/en-us/minitab/18/help-and-how-to/modeling-statistics/multivariate/how-to/principal-components/interpret-the-results/key-results/.

\bibitem{penn_umap}
J.~Hansen.
\newblock Umap.
\newblock https://www.math.upenn.edu/~jhansen/2018/05/04/UMAP/, May 2018.

\bibitem{umap_overview}
Adam~Pearce Andy~Coenen.
\newblock A dip into umap theory.
\newblock https://pair-code.github.io/understanding-umap/.

\end{thebibliography}

\end{document}